\documentclass[%
aip,
amsmath,amssymb,
reprint,%
floatfix
]{revtex4-2}

\usepackage{graphicx,array,booktabs,rotating}
\usepackage{dcolumn}
\usepackage{bm}
\usepackage{mathptmx}
\usepackage{multirow}
\usepackage{afterpage,float,color,xcolor}
\usepackage{hyperref}
\usepackage{etoolbox} 
\usepackage{tabularx,tabulary}
\usepackage[capitalize]{cleveref}
\usepackage{times}
\hypersetup{
	colorlinks,
	linkcolor={blue!100!black},
	citecolor={blue!100!black},
	urlcolor={blue!100!black}
}
\makeatletter
\newcommand\footnoteref[1]{\protected@xdef\@thefnmark{\ref{#1}}\@footnotemark}
\makeatother
\renewcommand{\figurename}{FIGURE}
\newcommand{\sk}[1] {\textcolor{black}{#1}}
\newcommand{\etal}{\textit{et al.}}
\usepackage[displaymath,pagewise]{lineno}

\begin{document}
	\preprint{AIP/123-QED}	
	\title[Phys. Fluids (2020) | Revised Manuscript (POF20-AR-03428)]{On the unsteady throttling dynamics and scaling analysis in a typical hypersonic inlet-isolator flow}
	\author{K. Raja Sekar}
	\affiliation{Department of Aeronautical Engineering, Bannari Amman Institute of Technology, Sathyamangalam-638401, India}%
	
	\author{S. K. Karthick}
	\email{skkarthick@ymail.com (Corresponding Author)}
	\affiliation{Faculty of Aerospace Engineering, Technion-Israel Institute of Technology, Haifa-3200003, Israel}%
	
	\author{S. Jegadheeswaran}
	\affiliation{Department of Mechanical Engineering, Bannari Amman Institute of Technology, Sathyamangalam-638401, India}%
	
	\author{R. Kannan}
	\affiliation{Department of Aerospace Engineering, Amrita School of Engineering, Amrita Vishwa Vidyapeetham, Coimbatore-641112, India}%
	
	\date{\today}
	
	\begin{abstract}
		The flow field in a two-dimensional three-ramp hypersonic mixed-compression inlet in a freestream Mach number of $M_\infty=5$ is numerically solved to understand the unsteady throttling dynamics. Throttling conditions are simulated by varying the exit area of the isolator in the form of plug insets. Different throttling ratios between $0\leq \zeta \leq 0.7$ in steps of 0.1 are considered. No unsteadiness is observed for $\zeta\leq 0.2$ and severe unsteadiness is found for $0.3 \leq \zeta \leq 0.7$. The frequency of unsteadiness ($f$) increases rapidly with $\zeta$. As $\zeta$ increases, the amount of reversed mass inside the isolator scales with the frequency and the exit mass flow rate. A general framework is attempted to scale the unsteady events based on the gathered knowledge from the numerical study. The inlet-isolator flow is modeled as an oscillating flow through a duct with known upstream design conditions like the freestream Mach number ($M_\infty$) and the isolator inlet Mach number ($M_i$). Factors like the mass occupied by the duct volume, the characteristic unsteady frequency, throttling ratio, and the exit mass flow rate through the duct are used to form a non-dimensional parameter $\beta$, which scales with the upstream design parameter $\xi=M_i/M_\infty$. The scaling parameters are further exploited to formulate a semi-empirical relation using the existing experimental results at different throttling ratios from the open literature.The unsteady frequencies from the present two-dimensional numerical exercise are also shown to agree with the proposed scaling and the resulting semi-empirical relation.
	\end{abstract}
	
	
	\maketitle
	
	\section{Introduction}\label{introduction}
	The starting characteristics of the hypersonic inlet are crucial for the operation and performance of the air-breathing hypersonic propulsion systems\cite{Raj2012,Bissinger1993,Murthy2001,Minucci1993,Marquart1991}. Hypersonic cruise vehicles, in general, have a higher service ceiling, typically in the range of 25-35 km, where the air density and pressure are thin\cite{Gnos1973,Iliff1995,Fan2011}. Inlets at such higher altitudes should be designed to achieve enough air compression for subsequent stable combustion\cite{Bogdonoff1999}. In such scenario, mixed compression inlets\cite{VanWie1994,VanWie1996} are preferred as the necessary compression is achieved in two different stages: 1. compression achieved through a series of weak oblique shocks emanating from the external ramps, and 2. a system of multiple reflecting shocks inside a sufficiently long duct called isolator\cite{Goldberg1971,Weir1989}. A longer external ramp produces a thicker boundary layer, and part of the flow into the isolator is distorted\cite{Fisher1986,Reddy1992}. Inside the isolator, multiple shock reflections or shock train/pseudo shocks\cite{Huang2016,Li2017,Li2018,Saravanan2020,Wang2020} form and further compress the flow. The isolator's flow passage suffers from severe shock wave boundary layer interactions\cite{Sriram2014,Sriram2016,Chen2019} (SWBLI) and becomes unsteady. The entire chain of shock systems is highly susceptible to fluctuations in the combustion chamber or the inlet's backpressure. The fluctuations associated with severe SWBLI separates the flow at hypersonic speed and renders the mass capturing event improbable. The resulting flow physics is unsteady, and the hypersonic inlet is considered unstart. Thus, the proper formation and response of shock wave systems in the external ramp and within the isolator to backpressure variations are important for the hypersonic inlet's efficient operation. 
	
	Inlet unstart introduces a low-frequency oscillatory flow phenomenon known as `inlet buzz.' A buzz\cite{Soltani2010,Hutzel2011,Chen2017,Im2018} is initiated when the ingested air mass at the supersonic speed in the inlet encounters a sudden rise in pressure at the isolator's exit. A series of wavefronts push the flow outside the inlet in order to accommodate the appropriate mass flow for the downstream conditions. As the flow readjusts, the incoming stream to the inlet remains supersonic and has no information of the downstream events. Thus the cycle continues in a periodical manner until the fluctuations subside. The resulting violent flow oscillation accompanied by the shock wave motions cause fatal damage to the airframe and even a loss of overall flight control. 
	
	Many researchers have devoted their time to identify the dominant frequency component producing the buzz in such hypersonic inlets. Hawkins and Marquart\cite{Hawkins1995} conducted time accurate pressure measurements in their experimental studies to predict the unstart characteristics of a two dimensional generic supersonic/hypersonic inlet. They attributed the back pressure-induced unstart to the instabilities developed in the diffused boundary layer. In the experiments of Rodi \etal \cite{Rodi1996}, time accurate pressure measurement in the dual-mode RAMJET/SCRAMJET inlet was conducted. The backpressure effect was simulated with the help of the throttling device at the isolator exit. Van Wie \etal \cite{VanWie1996} experimented with the start, unstart, and restart inlet characteristics at a freestream Mach number of $M_\infty=3$ in a rectangular inlet at different Reynolds number ($Re$) and cowl angles. They classified the choking induced unstart as hard unstart and large separation bubble induced unstart as soft unstart. The mechanism of the buzz cycle proposed by Tan \etal \cite{Tan2009} identified the flow spillage at the inlet entrance as the upstream source of instability. They concluded by proposing a feedback loop by combining the convection, the shock train motion, and the acoustic wave propagation.
	
	The experimental investigation by Wagner \etal \cite{Wagner2009} regarding the oscillatory behavior of the separation bubble has characterized the unstart conditions of the hypersonic inlet as non-oscillatory, low amplitude oscillatory, and high amplitude oscillatory using the downstream propagation of compression waves. In another experimental study by Wagner \etal \cite{Wagner2010} using PIV, the boundary separation in the isolator side wall was observed to initiate the unstart process. Srikant \etal \cite{Srikant2010} have examined the mechanism associated with the unstart process of the hypersonic inlet and investigated the method to mitigate the unstart process using active control strategies. Lee and Kang \cite{Lee2019} conducted a numerical study in a two-dimensional hypersonic inlet at $M_\infty=4.9$ to predict the effect of the boundary layer profile on the unstart phenomenon. They found that the low velocity prevailing in the inflow boundary layer profile decreases the pressure oscillation frequency and amplitude. The prediction accuracy of the frequency and amplitude of pressure oscillation depends on the boundary layer profile's accuracy. Saravanan \etal \cite{Saravanan2020} characterized the unstart and restart process of an isolator at $M_\infty=1.7$. They showed that the unstart mechanism is a continuous process with steady back pressure and a discontinuous process under oscillatory back pressure. A recent review by Chang \etal\cite{Chang2017} briefly discussed the unstart mechanism, detection, and control in a hypersonic inlet.
	
	Chen \etal \cite{Chen2017} experimented with a rectangular external compression supersonic inlet at $M_\infty=2.5$. They observed a regular reflection, Mach reflection, and $\lambda$-shaped pattern in the inlet flow by imposing throttling. They found that the intercommunication between the medium buzz and the big buzz governs the oscillatory flows at higher throttling ratios. Im and Do\cite{Im2018} presented a comprehensive review of the recent research where the flow unstart arising from the downstream flow choking had been studied. Berto \etal \cite{Berto2020} investigated experimentally the mechanism associated with inlet buzz in the hypersonic inlet at $M_\infty=5$. At the highest throttling ratio, the cowl lip normal shock was expelled, which produced a bow shock and led to inlet buzz. Recently, Wang \etal \cite{Wang2020} investigated the low-frequency unsteadiness in the isolator due to the background waves at $M_\infty=2.94$. They showed that the upstream mechanism exhibits a significant influence on the unthrottled flow field.
	
	\begin{figure}
		\centering{\includegraphics[width=1\columnwidth]{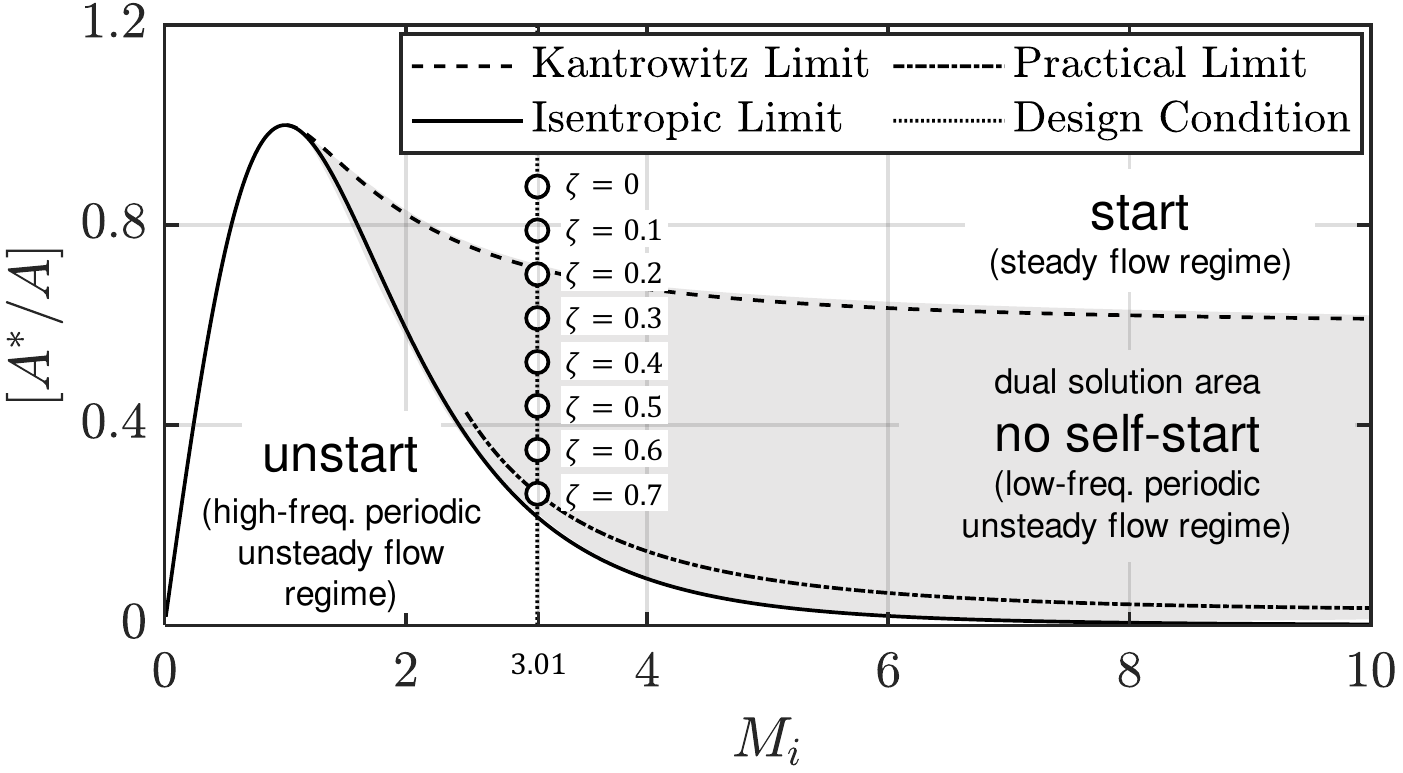}}
		\caption{A plot of ratio of generic throat area ($A^*$) and the capture area ($A$) to different inlet Mach numbers ($M_i$) to map out the regions of self-start, no self-start, and unstart zones with respect to the hypersonic inlet. The inlet Mach number to the considered hypersonic inlet in the present study is $M_i=3.1$, obtained from the two-dimensional $\theta-\beta-M$ relationships of compressible aerodynamics at the inlet cowl/ramp plane shown in Figure \ref{fig:model}. The throat area is taken as ramp height ($h^*$), and the capture area is taken as the constant height between the cowl tip and the ramp ($h_c=0.058$ m).}
		\label{fig:kant_limit}
	\end{figure}
	
	Most of those studies are targeted towards the understanding of unstart and a definitive mechanism to describe it. However, in reality, the inlet is subjected to many off-design conditions, which effectively changes the backpressure. In the work of Kantrowitz \cite{Kantrowitz} and Van Wie \etal \cite{VanWie1994,VanWie1996}, the existence of three regimes of hypersonic inlet flow operation is described: 1. steady flow regime where the inlet is considered as started, 2. unsteady flow regime where the inlet exhibits periodic flow oscillation, and it cannot be self-started, and 3. unstarted inlet with a steady or unsteady flow regime having aperiodic oscillations. These regimes are obtained by considering the inlet Mach number ($M_i$) and the generic capture area ratio to the throat area ($A^*/A$). The mapped out regimes are shown in Figure \ref{fig:kant_limit} for a better understanding. When the inlet operates in the dual solution regime, it produces periodic oscillation with a characteristic frequency ($f$), similar to buzz. When it enters the unstart regime, it oscillates at a higher frequency like a forward-facing cavity in the high-speed flows\cite{Engblom1995,Sudarshan2019}. 
	
	Interestingly, as mentioned in the review of Chang \etal\cite{Chang2017}, the oscillation patterns observed in the hypersonic inlet are different from the supersonic counterpart. Supersonic inlet buzz can be predicted using well-defined analytical solutions from the acoustic theory\cite{Trimpi1953,Mirels1955,Hankey1980,Newsome1984}. However, when the inlet Mach number is supersonic, mass filling up and local convection is strong. Hence, the traditional acoustic theory cannot predict $f$ for the oscillations observed in a hypersonic inlet.  Many systematic studies are carried out to generate hypersonic inlet buzz by carefully simulating the backpressure. The traditional throttling methods like the usage of plugs, as mentioned in the review of Im and Do\cite{Im2018} would be ideal for generating the downstream choking condition arising from the backpressure fluctuations. Experiments or simulations performed in that manner had produced different $f$ on different inlet configurations, and there was no universal scaling to relate $f$ for different throttling and operating conditions (like the freestream Mach number-$M_\infty$, inlet Mach number-$M_i$). An analytical solution or at least a semi-empirical relation would be beneficial to predict the buzzing frequency and design a suitable redundant airframe for hypersonic flight in the dual solution area as shown in Figure \ref{fig:kant_limit}.
	
	In the background of the motivation mentioned above, the following objectives are considered in the present paper:
	\begin{enumerate}
		\item{To perform a two-dimensional unsteady numerical simulation on a hypersonic inlet flow at a freestream Mach number of $M_\infty=$5 at different throttling ratios.}
		\item{To understand and characterize the evolving unsteady flow field at different throttling ratios by evaluating the thermodynamic and kinematic parameters.}
		\item{To obtain the spectral signature observed in the hypersonic inlet at different throttling ratios through spatio-temporal analysis.}
		\item{To identify the universal scaling variables required to formulate a semi-empirical relation that can relate the hypersonic inlet operating condition to the buzzing frequency using the existing experimental data in the open literature.} 
		\item{To use the computed frequency from the present two-dimensional numerical analysis to verify the formulated semi-empirical relation.}
	\end{enumerate}	
	
	The rest of the manuscript is organized as follows. The details about the numerical methodology, including the problem description, computational domain, meshing, solver, mesh independence, and validation, are given under various subheadings in Sec.\ref{sec:num_meth}. The evolution of unsteady events, variations in the performance parameters, and changes in the spectral signature for different throttling ratios are described in the subsections of Sec.\ref{sec:results}. The discussion on the scaling variables and the subsequent formation of a semi-empirical relation to identify the hypersonic buzz for the given design condition are given in Sec.\ref{sec:scaling}. The vital conclusions of the present study are given in the last section (Sec.\ref{sec:conclusions}).
	
	\begin{figure*}
		\centering{\includegraphics[width=0.9\textwidth]{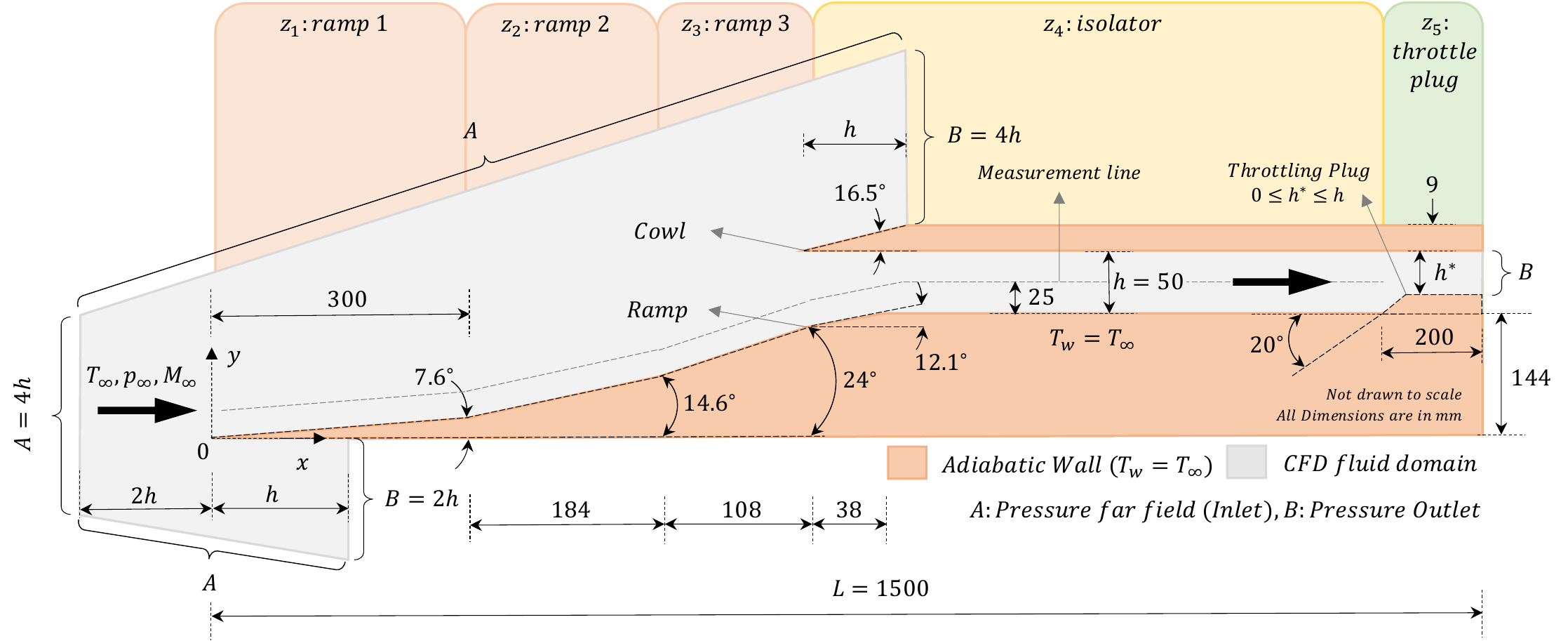}}
		\caption{A typical schematic showing the two-dimensional three-ramp hypersonic mixed-compression (compression achieved in three steps externally and through shock trains internally) inlet having a constant area isolator of considerable length with annotated geometrical values, computational domain limits, and boundary conditions (freestream conditions are given in Table \ref{table:flow_cond}). The different parts of the inlet are marked in the shaded color for spatial identification: ramps (pale-orange), isolator (pale-yellow), and throttle plug (pale-green). The sketch is not drawn to scale, and all the length dimensions are given in mm.}
		\label{fig:model}
	\end{figure*}
	
	
	
	
	\section{Numerical Methodology}\label{sec:num_meth}
	\subsection{Problem Description} \label{sec:prob_des}
	\begin{table}
		\caption{Freestream flow conditions achieved in the two-dimensional computational exercises at different throttling conditions\footnote{$\zeta=1-(h^*/h)$, where $\zeta$ is the throttling ratio varied as $0 \leq \zeta \leq 0.7$ in steps of 0.1, for example, $\zeta=0$ when $h^*=h$ and $\zeta=1$ when $h^*=0$. Here, $h^*$ is the minimum height available at the isolator exit after the insertion of plug during throttling, and $h$ is the isolator constant duct height.}.}
		\label{table:flow_cond}
		\begin{ruledtabular}
			\begin{tabular}{@{}ll@{}}    
				\textbf{Quantities} &
				\textbf{Values}\\ 
				\midrule
				Total Pressure ($p_0\times 10^5$, Pa) &
				2.96 \\
				Total Temperature ($T_0$, K) &
				1421.4\\
				Freestream Temperature ($T_\infty $, K) &
				236.9\\
				Freestream Pressure ($p_\infty$, Pa) &
				560\\
				Freestream Velocity ($u_\infty$, m/s) &
				1542.8\\
				Freestream Kinematic Viscosity ($\nu_\infty\times 10^{-3}$, m$^2$/s) &
				1.88  \\
				Freestream Density ($\rho_\infty \times 10^{-3}$, kg/m$^{3}$) &
				8.24\\
				Freestream Mach number ($M_\infty$) &
				5 \\
				Reynolds number ($Re \times 10^{6}=u_\infty/\nu_\infty$, m$^{-1}$)  &
				0.82 \\    			 
			\end{tabular}
		\end{ruledtabular}
	\end{table}
	
	A two-dimensional, rectangular, three-ramp, mixed-compression hypersonic inlet model, as shown in Figure \ref{fig:model} is taken to study the throttling dynamics. A generic hypersonic flight Mach number of $M_\infty=5$ at an altitude of 35 km is selected as an operational reference. Based on the flight altitude, the freestream conditions to the inlet are calculated and listed in Table \ref{table:flow_cond}. The overall inlet's length and the isolator's height are $L=1.5$ m and $h=0.05$ m, respectively. The shock-on-lip (SOL) design condition is achieved through three external compression ramps of having three different flow deflection angles ($\theta_1=7.6^\circ$, $\theta_2=7^\circ$, and $\theta_3=9.4^\circ$). The presence of three external compression ramps ensures the achievement of necessary flow compression of $p_i/p_\infty \approx$ 11 at the considered flight altitude and an inlet Mach number of $M_i=3.09$. Value of $p_i/p_\infty$ and $M_i$ are calculated using oblique shock relations ($\theta-\beta-M$)\cite{Zucrow1976} at the cowl/ramp plane near the isolator entrance. Further compression is realized through the successive reflected shocks/shock-train inside the isolator portion of the hypersonic inlet of length $\mathcal{L}=0.708$ m. The rest of the current inlet's geometrical features are adopted from some of the authors' previous article \cite{Rajasekar2019}.
	
	The inlet's unsteady characteristics arise from the combustion chamber back pressure fluctuations or drop in vehicle velocity, which unstart the inlet. Combustion chamber pressure variations reduce the air mass flow rate into it, and during the throttling process, such changes are inevitable. In the present simulation, throttling is achieved using a wedge ramp (plug) at the isolator's exit, as shown in Figure \ref{fig:model}. The wedge ramp reduces the isolator height from $h$ to $h^*$, thereby alters the net mass flow rate that is allowed into the combustion chamber and throttles the vehicle. Owing to the two-dimensional inlet model, $[A^*/A]=[h^*/h]$, as the lateral reference dimension is taken as 1 m. A parameter called throttling ratio ($\zeta$) is thus defined to quantify the extent of throttling as,
	\begin{linenomath} 
	\begin{equation}
	\zeta = 1 - \left(\frac{A^*}{A}\right) = 1 - \left(\frac{h^*}{h}\right).
	\label{eq:tr}
	\end{equation}
	\end{linenomath}
		
	The initial predictions of possible throttling ratios are made using the plots of isentropic area ratio variation and Kantrowitz\cite{Kantrowitz} limit for different inlet Mach numbers ($M_i$) as shown in Figure \ref{fig:kant_limit}. The lower limit of the throttling ratio ($\zeta=0$) indicates the started flow with the design mass flow rate passing into it. On the other hand, the upper limit comes from the practical limit of maximum area contraction ratio \cite{Murthy2001} for which the inlet could still be started ($\zeta=0.7$) as shown in Figure \ref{fig:kant_limit}. Larger values render the inlet into the unstart regime of operation. Hence, for the present analysis, the throttling ratio is varied only between $0 \leq \zeta \leq 0.7$ in steps of 0.1.
	
	\subsection{Computational Domain, Meshing, and Flow Solver} \label{sec:comp}
	The hypersonic inlet is enclosed in a partial external and internal computational fluid domain, shown as shaded gray color in Figure \ref{fig:model}. The upstream and downstream domain length in the frontal portion about the origin is $2h$ and $h$. The terminal fluid domain behind the cowl's external portion is away from the cowl tip by a length of $h$. The vertical length of the computational domain enclosing the frontal portion of the pressure far-field boundary is of dimension $4h$. Similarly, the pressure outlet boundary's vertical length, external to the cowl, and the ramp are kept as $4h$ and $2h$, respectively. The internal fluid domain extends from the cowl to the isolator exit. An appropriate number of individual meshes are prepared for different throttling ratios ($\zeta$) based upon the different solid ramp height-$h^*$ at the isolator's exit. Pressure far-field inlet boundary conditions are used at locations marked as $A$, and pressure outlet conditions are applied at zones marked as $B$ in Figure \ref{fig:model}. The edges surrounding the fluid domain marked in pale-orange color are declared as an adiabatic wall.
	
	A compact domain and an appropriate meshing strategy reduce the overall mesh counts and save the computational time required to resolve the flow. The fluid dynamics inside the inlet is of primary importance in the present study. As the flow is in the hypersonic regime, the fluid domain length is kept as minimum as possible. A commercial meshing software package from Ansys-ICEM\textsuperscript{\tiny\textregistered} is utilized to prepare the computational grids. A structured meshing scheme is adopted, and the computational fluid domain is packed with quadrilateral cells. The equisize skewness parameter is maintained at 0.2 for 96\% of the cells. The turbulence wall parameter ($y^+$) is kept less than one to resolve the boundary layer effects. The progression of mesh cell spacing in the isolator is not kept more than 1.2 in both the streamwise and transverse directions for the fine grid ($1.1\times 10^5$ cells) adopted in the study. The details of the mesh cell counts and the mesh independence study are given in Sec.\ref{sec:grid}.
	
	Numerical analysis is performed using a commercial computational fluid dynamics (CFD) package from Ansys-Fluent\textsuperscript{\tiny\textregistered}. The solver discretizes the fluid domain based on the finite volume schemes\cite{Ansys}. The unsteady Favre/Reynolds averaged Navier-Stokes (URANS) equations (comprising of continuity, momentum, energy, and the respective equations to approximate and close the turbulent equations) are solved using the coupled-pressure based solver with compressibility corrections. The fluid turbulence is modeled using a shear stress transport\cite{Menter1993,Menter1994} based two-equation eddy viscosity model called as SST-$k\omega$ which is known to predict the hypersonic inlet flow features as seen in the experiments\cite{Roy2006,Oliden2013,Wang2013,Deng2017,Lee2019}. The validation of the chosen turbulence model with experiments in hypersonic inlet flow is discussed in Sec.\ref{sec:grid}. The flow field is solved with air as the ideal gas, and the fluid's viscosity is computed through Sutherland's three equation model. All the flow equations are discretized both spatially (explicit) and temporally (implicit) with second-order accuracy. The gradients are resolved using Green-Gauss node based techniques, and a hybrid initialization approach is adopted for rapid convergence during the steady-state simulation. Later, unsteady flow solutions are sought at a fine time step of $[\Delta t/T]=1 \times 10^{-3}$ (where, $T= 1$ ms) with 20 iterations per unit step. Justification on the selected time step size is provided in Sec.\ref{sec:grid}. An absolute convergence criterion of $10^{-5}$ is achieved at every time step in the continuity equation's scaled-residuals.     
	
	\subsection{Mesh Independence, Time Dependence, and Solver Validation} \label{sec:grid}
	
	\begin{figure}
		\centering{\includegraphics[width=\columnwidth]{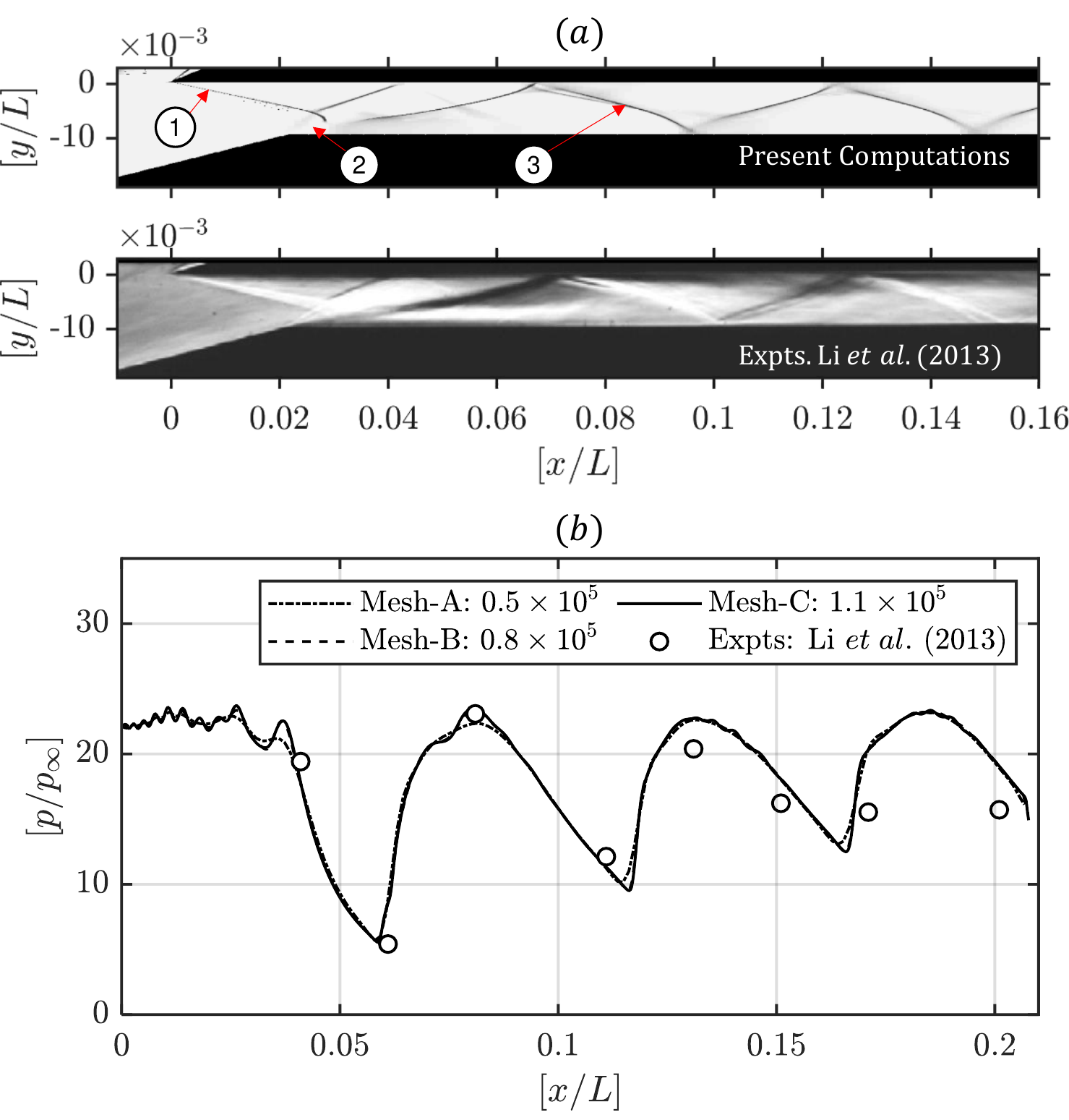}}
		\caption{(a) Qualitative spatial comparisons of the shock pattern formations in the isolator section of the hypersonic inlet between the numerical shadowgraph (Mesh \textit{C}) and the experimental schlieren\cite{Li2013}. Key flow features: 1. Shock emanating from the cowl, 2. Shock wave boundary interaction zone, 3. Reflected shocks inside the isolator; (b) Graph showing the computed wall-static pressure variations from the RANS based simulation on the cowl-side portion of the isolator along the $x$-direction ($L=1$ m, reference length considered for this particular analysis) for three different mesh densities (Mesh \textit{A},\textit{B}, and \textit{C}). The computed wall-static pressure measurements are also shown to be comparable with the experimental findings of Li \etal \cite{Li2013} (Image reproduced with permission from the authors of J. AIAA., 51(10), 2013, Copyright 2013).}
		\label{fig:mesh_val}
	\end{figure}
	
	Two-dimensional structured meshes are made at three different mesh densities: Mesh-A (coarse, $0.5 \times 10^5$ cells), Mesh-B (medium, $0.8 \times 10^5$ cells), and Mesh-C (fine, $1.1 \times 10^5$ cells), as mentioned in Sec.\ref{sec:comp} to find the dependence of mesh density on the final numerical solution and also to validate the adopted solver. Experiments of Li \etal \cite{Li2013} are considered for the mesh independence and solver validation exercises. A qualitative comparison is made between the time-averaged experimental schlieren and the steady-state numerical shadowgraph from Mesh-C in Figure \ref{fig:mesh_val}, where the shock reflection patterns arising from the cowl tip, the shock-wave boundary-layer interactions on the ramp wall, and reflecting shock waves inside the isolator are shown to be spatially similar. For quantitative comparisons, experimental wall-static pressure measurements ($p/p_\infty$) obtained on the cowl side (top side) of the inlet are compared with the numerical results from all the three different meshes. The experimental data points closely match the results from the Fine mesh (Mesh-C) among the others. The deviations are attributed to the sensor head size and three-dimensional effects present in the actual experiments.
	\begin{figure}
		\centering{\includegraphics[width=\columnwidth]{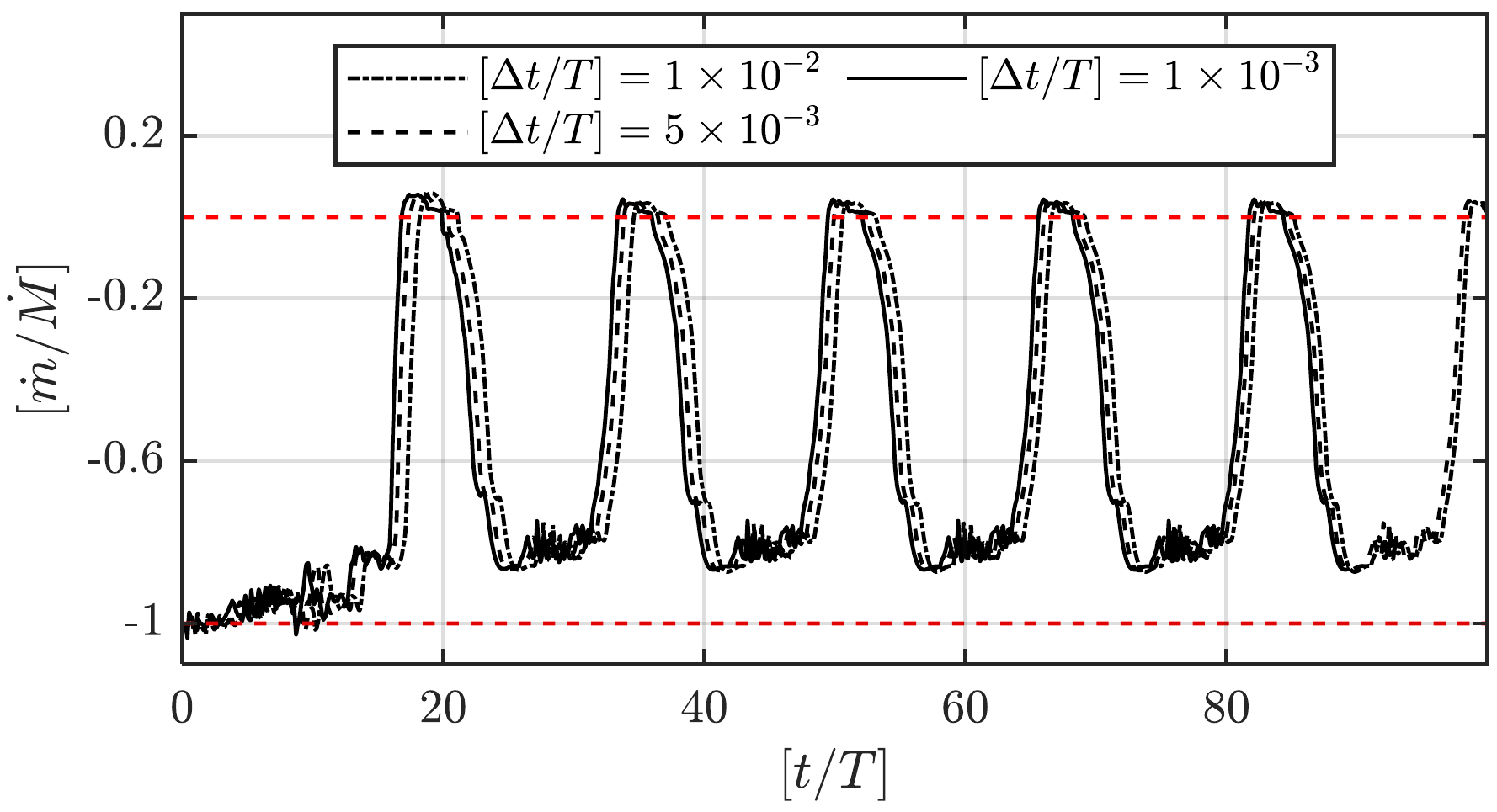}}
		\caption{Graph showing the exit-mass flow rate variations at $[x/L]\sim 0.7$ inside the isolator section of the hypersonic inlet at different temporal resolutions for $\zeta=0.3$ (note: variations are on the negative scale as the mass flux is coming out of the measurement line/plane). Non-dimensionalizing parameter $\dot{M}$ represents a steady reference mass flow rate of magnitude 2.45 kg/s achieved for $\zeta=0$, and $T$ is an arbitrarily chosen reference time of 1 ms. Dotted red-lines at $[\dot{m}/\dot{M}]=-1$ and $[\dot{m}/\dot{M}]=0$ mark the reference boundary, where the values of $[\dot{m}/\dot{M}]>0$ signify reverse flow in the isolator.}
		\label{fig:time_dep}
	\end{figure}
	
	A similar numerical exercise is performed using the fine mesh (Mesh-C) for unsteady flow simulation on the domain mentioned in Figure \ref{fig:model}. Flow unsteadiness due to throttling is expected for $\zeta \geq 0.3$ from the simple plot made in Figure \ref{fig:kant_limit} using the Kantrowitz limit and isentropic limit for a designed inlet Mach number of $M_i=3.09$. Three different time steps: $[\Delta t/T]=1 \times 10^{-2}$, $[\Delta t/T]=5 \times 10^{-3}$, and $[\Delta t/T]=1 \times 10^{-3}$ are considered for the particular case of $\zeta=0.3$ to see the influence of time step dependence. The exit-mass flow rate at the end of the isolator is monitored, as shown in Figure \ref{fig:time_dep}. In general, the solution is obtained until $[t/T]=100$ to capture at least five unsteady cycles. A reference mass flow rate of 2.45 kg/s obtained at $\zeta=0$ (where the flow is steady) is used to non-dimensionalize the plot about the $y$-axis. The negative values represent the efflux (flow out of the exit plane), and the positive values show the reversed flow into the isolator. The dynamic events are captured in detail for the smallest time step ($\Delta t/T=1 \times 10^{-3}$).
	
	\begin{figure}
		\centering{\includegraphics[width=\columnwidth]{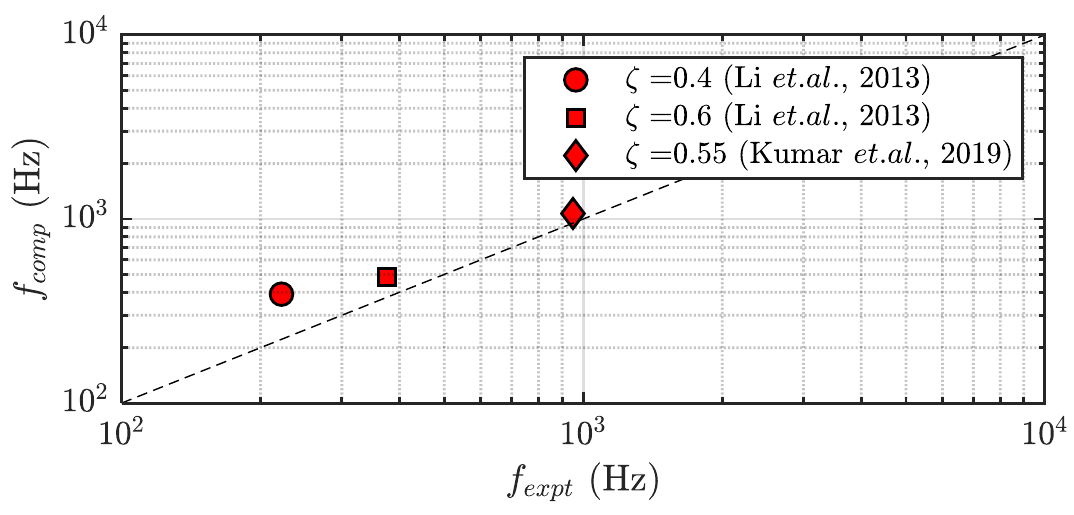}}
		\caption{Validation of dominant unsteady frequency ($f$) between the two-dimensional URANS simulations and the existing experiments (Li \etal \cite{Li2013} and Kumar \etal \cite{Devaraj2020}) from the open literature at different throttling ratios ($\zeta=$0.4\cite{Li2013}, 0.6\cite{Li2013}, and 0.55\cite{Devaraj2020}).}
		\label{fig:freq_validation}
	\end{figure}
	
	Experimental data of Li \etal \cite{Li2013} and Kumar \etal \cite{Devaraj2020} at different throttling ratios ($\zeta=0.4,0.55,$ and $0.6$) are used to further validate the fidelity of the present solver in predicting the unsteady features. The meshes are made using the respective geometries given in the literature \cite{Li2013,Devaraj2020}, and the throttling is achieved using the plug as described for the present model. The dominant unsteady frequency from those experiments are compared with the respective numerical simulation as part of the solver validation exercise on the fine mesh and at $[\Delta t/T]=1 \times 10^{-3}$. The comparative values are plotted in Figure \ref{fig:freq_validation}. The computed values from the URANS simulations closely match the experiments, and the deviations are attributed to the three-dimensional effects. A separate section (see Sec.\ref{sec:scaling}) at the end of the paper discuss deviations from three-dimensional effects and the existence of universal scaling variable in detail. Thus, for the rest of the present paper's analysis, solutions from a fine mesh with $[\Delta t/T]=1 \times 10^{-3}$ are only considered.
	\begin{figure*}
		\centering{\includegraphics[width=\textwidth]{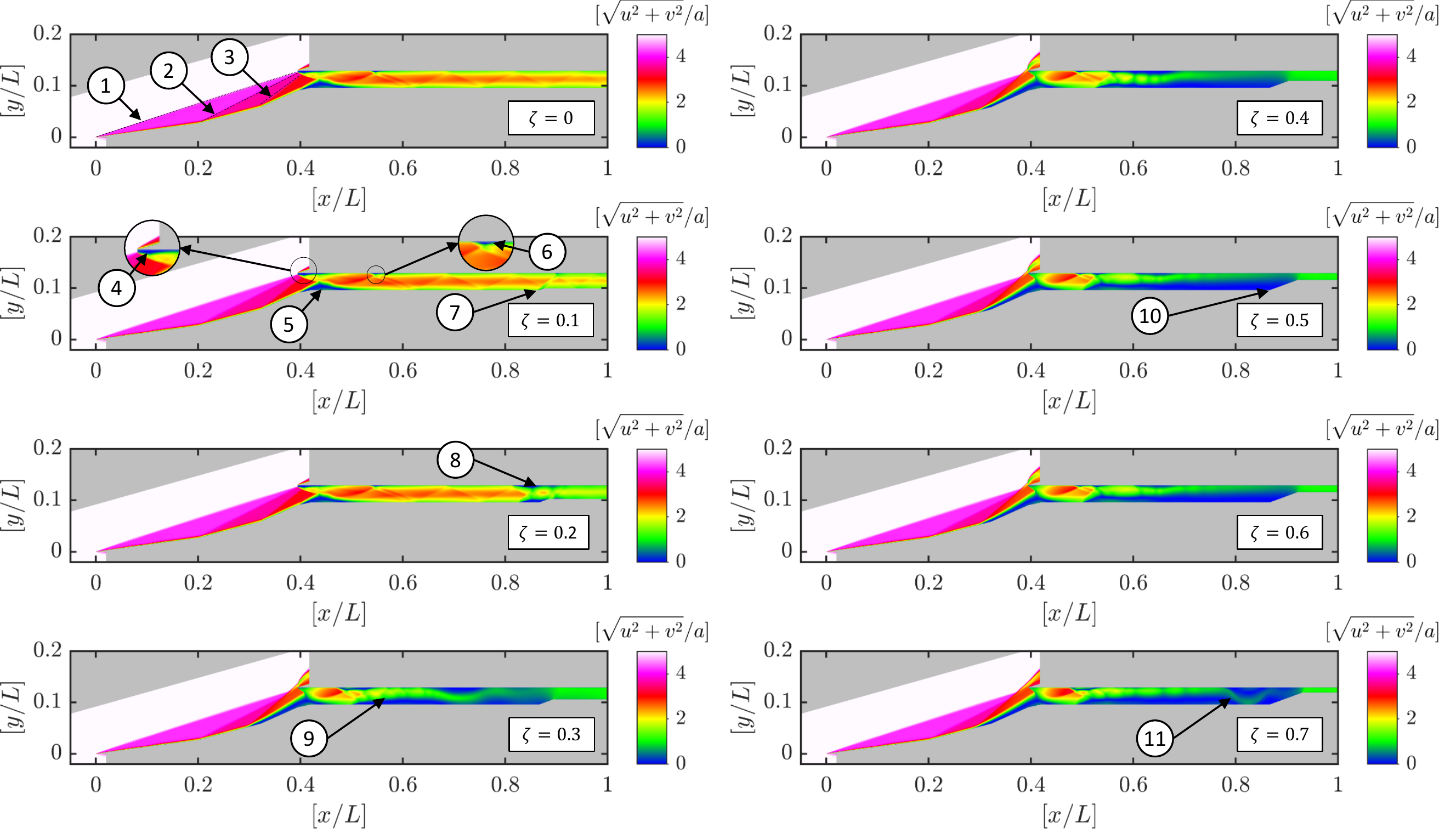}}
		\caption{Typical instantaneous contour plots (\href{https://youtu.be/LGo_uGc24zg}{Multimedia View}) of Mach number observed at an arbitrary time step during the simulation of different throttling ratios. A steady flow field is observed up to $0 \leq \zeta \leq 0.2$ and an unsteady flow field is seen for $0.3 \leq \zeta \leq 0.7$. Key flow features: 1. shock from ramp-1, 2. shock from ramp-2, 3. shock from ramp-3, 4. shock wave boundary layer interaction (SWBLI) on the cowl wall, 5. SWBLI on the ramp wall, 6. secondary SWBLI inside the isolator, 7. throttling plug induced shock, 8. choking induced by the throttling plug, 9. formation of unsteady pseudo shock/shock train, 10. increasing throttling plug height, 11. the sinuous upstream motion of the pseudo shock/shock train.}
		\label{fig:diff_zeta}
	\end{figure*}
	\begin{figure*}
		\centering{\includegraphics[width=\textwidth]{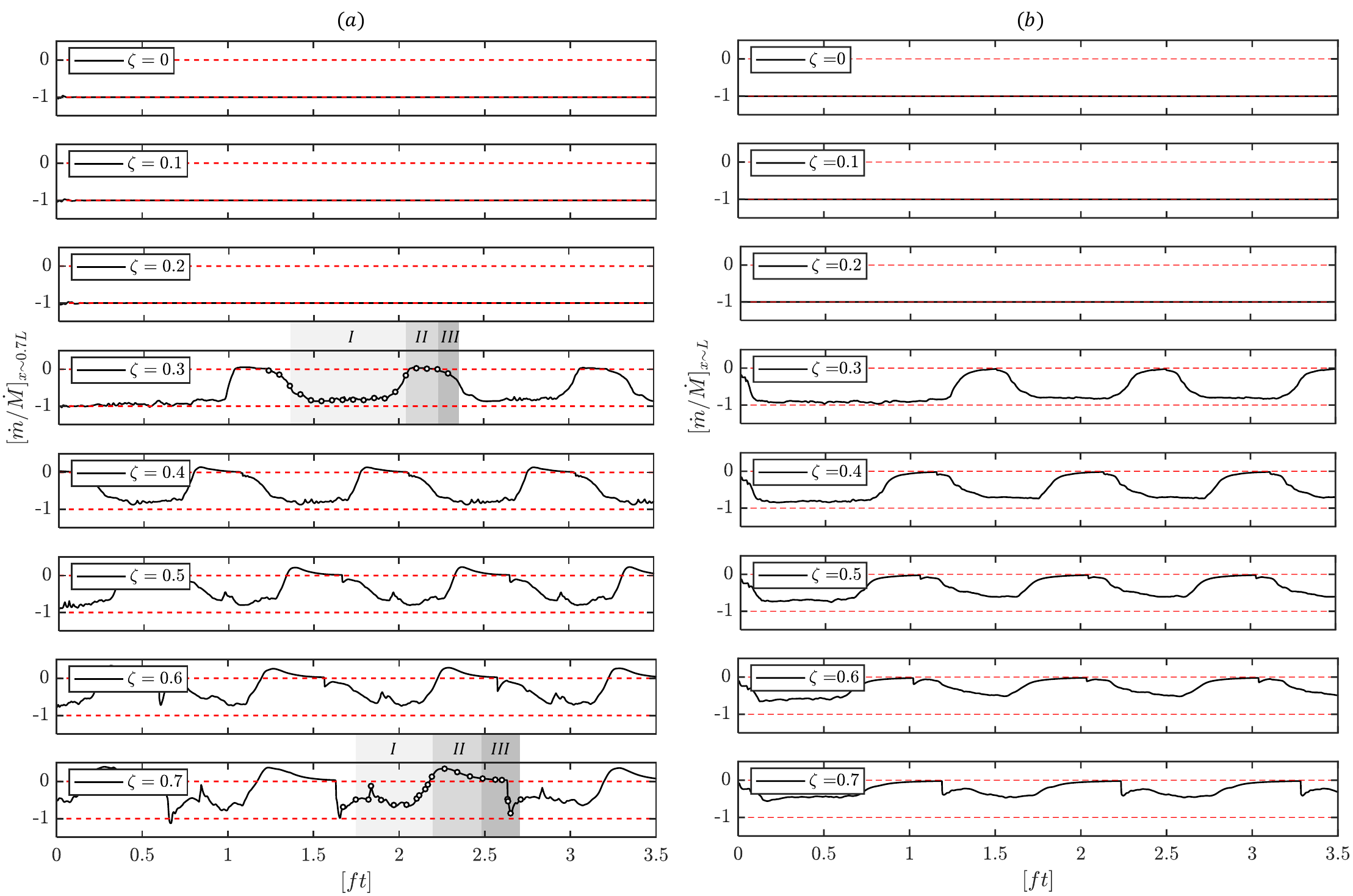}}
		\caption{Variation of non-dimensionalized exit mass flow rate ($\dot{m}/\dot{M}$) over the course of non-dimensionalized simulation time ($ft$) for different throttling ratio ($0 \leq \zeta \leq 0.7$) highlighting the time-period of the encountered unsteady cycle at two different stations: (a) $[x/L]\sim 0.7$ and (b) $[x/L]\sim 1$. The dotted red line at $[\dot{m}/\dot{M}]=0$ marks the boundary where the flow is reversed as $[\dot{m}/\dot{M}]>0$. The red dotted line at $[\dot{m}/\dot{M}]=1$ marks the maximum non-dimensionalized exit mass flow rate encountered in the present study, which is the steady mass flow rate observed for the throttling condition $\zeta=0$. The filled circle markers are given as a reference for a particular region of $[ft]$ at $\zeta=0.3$ and $\zeta=0.7$ in (a) whose instantaneous flow field can be seen in Figure \ref{fig:diff_zeta}. Three different stages encountered during a typical unsteady cycle are marked: I. mass-filling up the isolator duct, II. back pressure propagating upstream, and III. disgorging of shock systems.}
		\label{fig:mass_TR}
	\end{figure*}
	\begin{figure*}
		\centering{\includegraphics[width=\textwidth]{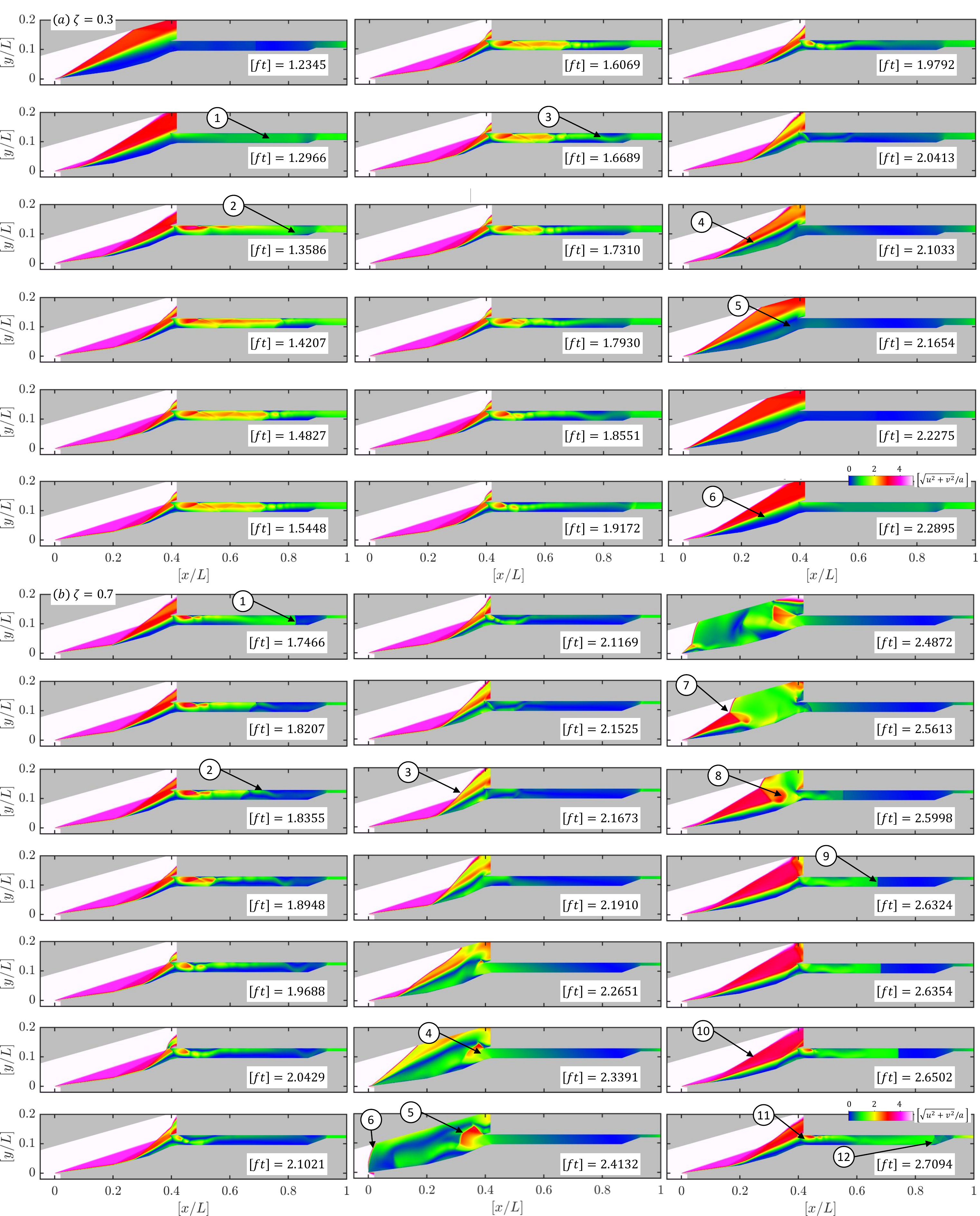}}
		\caption{Instantaneous contour plots of Mach number at different time instants in an unsteady cycle for (a) $\zeta=0.3$ and (b) $\zeta=0.7$ as marked in Figure \ref{fig:mass_TR}-a. Key flow features in (a): 1. mass filling-up the isolator, 2. receding shock system, 3. transformation to sinuous pseudo-shock cells, 4. disgorged shock system, 5. subsonic flow spillage, 6. collapsing shock system. Key flow features in (b): 1. receding normal shock, 2. sinuous pseudo-shocks, 3. disgorged shock system (inflation), 4. supersonic spillage, 5. Mach disc from the supersonic jet, 6. curved shock in the leading edge, 7. triple-point of the collapsing shock system, 8. vortex shedding from triple-point, 9. downstream propagating normal shock, 10. collapsing shocks, 11. larger-separation bubble blocking the inlet, 12. multiple shock reflections near the throttling plug.}
		\label{fig:comb_cycle}
	\end{figure*}
	\section{Results and Discussions} \label{sec:results}
	
	The numerical results are discussed in three fronts: 1.The evolution of the unsteady flow field from throttling. 2. The variations observed on the performance parameters of the inlet due to throttling, 3. The spectral signature due to the resulting unsteady events observed in the inlet at the time of throttling through the $x-t$ plot and fast-Fourier transform (FFT) analysis. The existence of universal scaling to predict the dominant frequency of unsteadiness during throttling and demonstrating the usefulness of simplified two-dimensional CFD analysis with corrections to get the frequency right is shown in the last section (Sec.\ref{sec:scaling}) before vital conclusions.
	
	An instantaneous contour plot of Mach number defined through the ratio of velocity magnitude and free stream sound velocity ($\sqrt{u^2+v^2}/a$) is shown in Figure \ref{fig:diff_zeta}. The contour plot depicts the unfolding events observed in a mixed-compression hypersonic inlet at different throttling ratios ($\zeta$). \sk{From the statistics of flow kinematics and the mass flow rate plots shown later in the discussions (Figure \ref{fig:diff_zeta} and Figure \ref{fig:mass_TR}), cases of $0 \leq \zeta \leq 0.2$ are observed to be steady}. The isolator's steady flow field consists of reflecting shocks between the cowl and ramp wall of the isolator. Firstly, the cowl shock impinges on the ramp wall's shoulder and results in the flow separation due to the shock-wave boundary layer interaction (SWBLI) and leads to the formation of a separation bubble of considerable thickness (almost half of the isolator section height, $h$). The induced separation shock from the separation bubble further hits the cowl wall and results in the secondary separation bubble formation. The secondary separation bubble develops a weak separation shock and the secondary shock reflection. Both the weak separation shock and the secondary reflection shock merge and impinge and reflect within the isolator walls. The eventual interactions produce \sk{multiple shock reflections} until the end of the isolator exit, and the flow at the outlet pressure boundary remains partially supersonic. All the key flow features during the steady-state operation are marked in Figure \ref{fig:diff_zeta}.
	
	On the other hand, cases between $0.3 \leq \zeta \leq 0.7$ demonstrate a state of unsteadiness, which is periodic. The frequency of dominant periodic fluctuations progressively increases as $\zeta$ varies between 0.3 and 0.7. In the contour plots of Figure \ref{fig:diff_zeta} (\href{https://youtu.be/LGo_uGc24zg}{Multimedia View}), the presence of sinuous tail at the end of the pseudo-shock train in the isolator section for $0.3 \leq \zeta \leq 0.7$ shows the ongoing unsteady events unfolding in the hypersonic inlet. \sk{The sinuous motion of the pseudo-shock-train shown in Figure \ref{fig:diff_zeta} is analogous to the flapping sonic or supersonic free\cite{Chaudhary2020,Rao2020} or confined jet\cite{Karthick2016,Karthick2017}. The receding flow in the isolator separates due to strong shocks. The resulting flow column behind the shock wave becomes unstable and flap, leading to the formation of the sinuous tail.} The occurrence of unsteadiness at particular $\zeta$ are consistent with the predicted zones of unsteadiness (or where the inlet cannot self start) from the $[A^*/A]$ and $M_i$ plot shown in Figure \ref{fig:kant_limit}. This unsteadiness alters the net mass flow rate into the isolator, thereby compromising the inlet's performance parameters.
	
	\subsection{On the evolution of unsteady events} \label{sec:unsteady_events}
	The severity of the flow unsteadiness is observed to increase with $\zeta$, as shown through the instantaneous contour plots in Figure \ref{fig:comb_cycle}. In order to evaluate the unsteadiness intensity, mass flow rates at two \sk{arbitrary} stations ($x/L \sim 0.7$ and $x/L \sim 1$) at the isolator section are monitored and plotted in Figure \ref{fig:mass_TR} for $\zeta$ varies. The plot immediately shows that for cases of $0 \leq \zeta \leq 0.2$ remains fairly steady and cases of $0.3 \leq \zeta \leq 0.7$ exhibits a periodic unsteadiness. Mass flow rate through the inlet for $0 \leq \zeta \leq 0.2$ remains the same at both the station and the magnitude of mass flow rate in specific at $\zeta=0$ ($\dot{m}=2.45$ kg/s) is used for non-dimensionalizing the mass flow rates for the other cases $\zeta$. The $x$-axis is non-dimensionalized using the product of local \sk{time ($t$, s)} and the dominant unsteady low-frequency component ($f$, Hz) to compare the basic features between different $\zeta$. The station at $[x/L]\sim 0.7$ inside the isolator reveals better insights into the flow events than the isolator exit ($x/L \sim 1$). \sk{The back pressure-induced upstream propagation of the shock system} and the resulting reversed mass flow rate could be easily monitored at $[x/L]\sim 0.7$. However, in each of the stations, the decreasing mass flow rate as $\zeta$ increases from 0.3 to 0.7 is evident. There are many sharp peaks and valleys as $\zeta$ increases, indicating the vigorous transit of shocks and expansion waves across the isolator. For further discussions, the results at $[x/L] \sim 0.7$ inside the isolator are considered.
	
	Two dotted-red lines are marked in the graphs shown in Figure \ref{fig:mass_TR} as references. Line passing through $[\dot{m}/\dot{M}]=-1$ represents the maximum efflux (going out the measurement station) from the isolator section with respect to $\zeta=0$ and the line passing through $[\dot{m}/\dot{M}]=0$ demarcates the streamwise flow region ($\dot{m}/\dot{M}<0$) from the reverse flow region ($\dot{m}/\dot{M}>0$). From Figure \ref{fig:mass_TR}a, the amount of time that the inlet experience the reverse flow increases gradually. \sk{The intersecting area on the positive region of Figure \ref{fig:mass_TR}a ($\dot{m}/\dot{M}>0$) between the curves of the non-dimensionalized mass flow rate ($\dot{m}/\dot{M}$) and the dividing red-dotted line at $\dot{m}/\dot{M}=0$ for a unsteady single cycle is the reversed flow mass.} The extent of the reverse flow experienced by the isolator indeed alters the flow events inside it. The unfolding unsteadiness in a typical cycle is classified into three stages\cite{Chang2017}: I. filling-up the isolator with the captured air-mass from the freestream, II. back pressure-induced choking creating upstream propagating waves, and III. disgorging of the shock systems inside and outside the isolator. The respective stages are shown in Figure \ref{fig:mass_TR}a for an unsteady cycle at $\zeta$=0.3 and 0.7. These stages are also shown in Figure \ref{fig:comb_cycle} through the instantaneous contour plot of Mach number and the corresponding time instants for reference.
	
	As $\zeta$ increases in Figure \ref{fig:mass_TR}a from 0.3 to 0.7, the dominant frequency content increases, and the successive cycles are packed densely. At $\zeta=0.3$, the wedge ramp at the exit of the isolator exit produces a sufficiently larger flow passage compared to $\zeta=0.7$. Hence the associated total pressure loss across the wedge ramp shock is minimal for $\zeta=0.3$. Consequently, to accommodate the mass flow rate corresponding to the new stagnation conditions established at the isolator exit, a series of wavefronts start to travel upstream. These wavefronts perturb the existing shock systems and create sinuous pseudo-shock cells in the isolator, which are seen for both the cases ($\zeta=0.3$ and 0.7). However, the upstream speed is comparatively smaller for $\zeta=0.3$. As the pseudo-shock cells are pushed into the inlet's entrance, subsonic reverse flow is established in the isolator. Just like the `inflate' phase of the unsteady shock motion in the axisymmetric spiked body flows\cite{Feszty2004,Sahoo2016,Sahoo2019}, the receding flow in the isolator pushes the upstream shock system or disgorge to give way to the exiting reverse flow against the stream. For $\zeta=0.3$, the exiting flow is subsonic (or subsonic spillage) due to the lower total pressure loss ($ft=2.22$ in Figure \ref{fig:comb_cycle}a), whereas for $\zeta=0.7$ the expelled flow behaves like a supersonic jet\cite{EdgingtonMitchell2014,Rao2020} or counterflow jet\cite{Sharma2020,Desai2020}, causing supersonic spillage. Some of the under-expanding jet characteristics like the expansion fans, Mach disc, and shock barrel are visibly seen during the expulsion ($ft=2.41$ in Figure \ref{fig:comb_cycle}b).
	
	The shock systems are pushed to the leading edge of the inlet and disgorged extensively for $\zeta=0.7$, just like the `with-hold' phase of the unsteady spiked body flows\cite{Feszty2004,Sahoo2016,Sahoo2019} until the mass inside the isolator is expelled or spilled. The standing shock systems in the leading edge of the hypersonic inlet collapse immediately, just like the `collapse' phase of the unsteady spiked body flows\cite{Feszty2004,Sahoo2016,Sahoo2019} at the end of the fluid expulsion stage from the isolator. The distinct characteristics of the `collapse' phase at $\zeta=0.7$ involve the formation of strong vortices. They are shed from the triple-point of the resulting shock interactions on the hypersonic inlet's external ramp. The resulting flow establishes the design supersonic inlet Mach number ($M_i=3.09$), which sends the same amount of mass flow rate into the isolator as before.
	
	The supersonic flow into the isolator is seen through the travelling normal shock into the isolator from the entrance between $2.59 \leq [ft] \leq 2.65$ in Figure \ref{fig:comb_cycle}b. Once it encounters the smaller flow passage at the isolator's exit, necessary total pressure loss occurs, and thus, the events reoccur in a self-sustained manner as similar to the previous cycle. However,  the shock systems are not pushed to the leading edge and quickly collapse owing to the subsonic jet expulsion for $\zeta=0.3$. The shock systems are not disgorged as extensively as it is at $\zeta=0.7$. The mass recharging process inside the isolator by the incoming flow is quicker as shown in Figure \ref{fig:comb_cycle}a between $1.23 \leq [ft] \leq 1.35$. The longer time taken for the upstream traveling wavefronts ($1.35 \leq [ft] \leq 2.04$), immediate collapse, and quicker recharge of the isolator by the incoming flow is the primary reason for the production of comparatively low-frequency periodic unsteadiness at lower throttling ratio ($\zeta=0.3$) than at $\zeta=0.7$.
	
\begin{figure*}
		\centering{\includegraphics[width=0.7\textwidth]{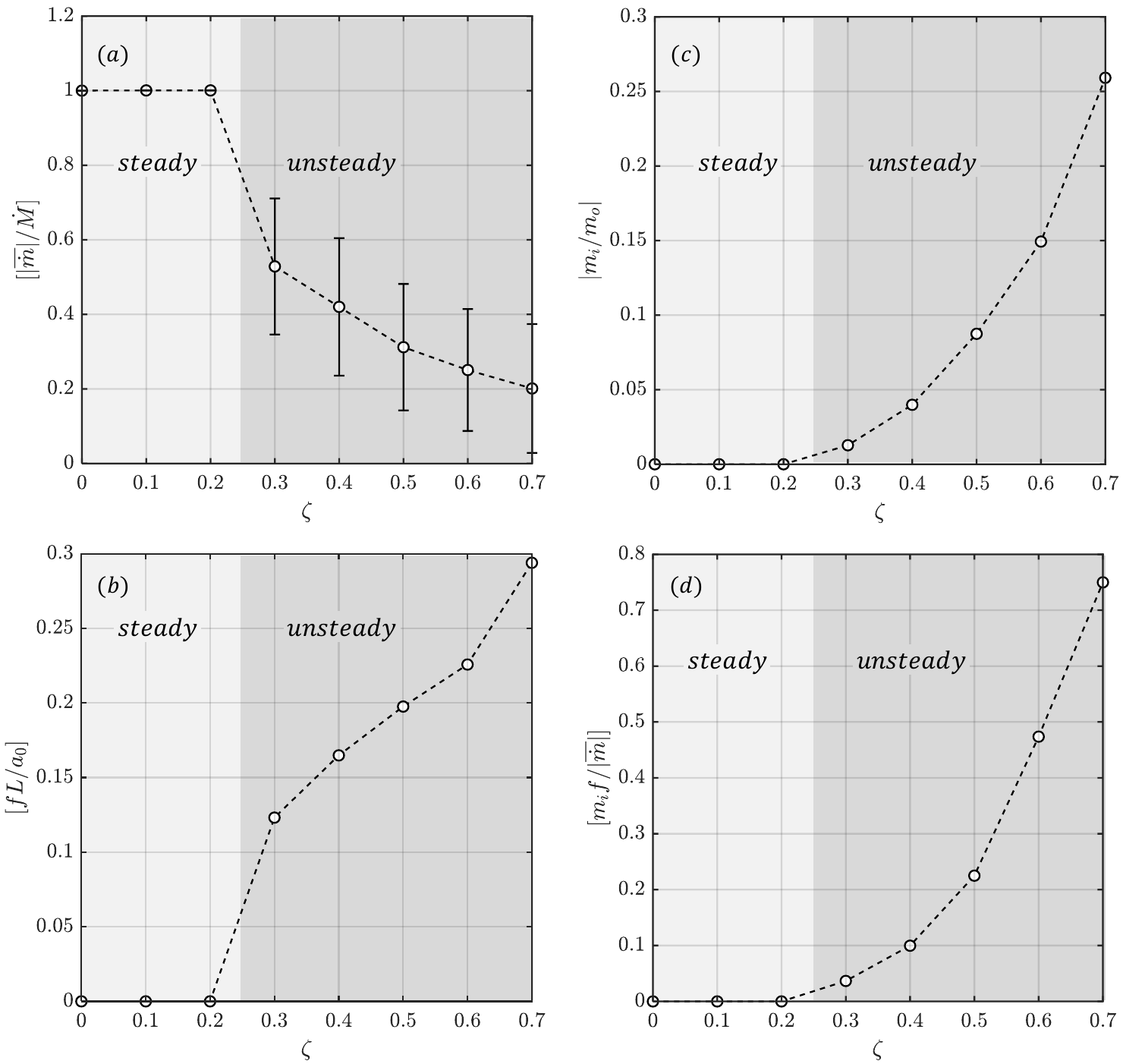}}
		\caption{Scaling relations arising from the mass flow rate, reversed mass flow, and the characteristic frequency of unsteadiness for different throttling conditions ($\zeta$): (a) Non-dimensionalized magnitude of the time-averaged mass flow rate observed at the isolator outlet ($[x/L]\sim 1$). \sk{Error bars correspond to the standard deviation of the cyclic variations in the mass flow rate for $0 \leq \zeta \leq 0.7$ as shown in Figure \ref{fig:mass_TR}b}; (b) Non-dimensionalized variation of the dominant unsteady frequency obtained from the fast Fourier transform of the unsteady mass flow rate signals at $[x/L]\sim 1$; (c) Variation of the magnitude ratio between the in-coming and out-going mass observed at a reference location of $[x/L]\sim 0.7$; (d) Variation of the scaled frequency parameter using the reversed mass and average mass flow rate variables.}
		\label{fig:mass_freq_TR}
	\end{figure*}

	The hypersonic inlet's mass flow rate plays an important role in identifying the unsteady events inside the isolator. A typical plot of the non-dimensionalized mass flow rate exiting through the pressure outlet boundary is plotted in Figure \ref{fig:mass_freq_TR}a for different throttling ratios ($\zeta$). As $\zeta$ increases until the Kantrowitz limit ($0 < \zeta < 0.3$) shown in Figure \ref{fig:kant_limit}, values of $[\dot{m}/\dot{M}]$ remains constant and the flow remains steady. Once $\zeta$ falls within the Kantrowitz limit and the practical limit of area contraction in the isolator ($0.3 \leq \zeta \leq 0.7$), flow unsteadiness is evident. The inlet enters into the periodically oscillating `no self-start' regime where dual solution exist\cite{Kantrowitz}. A drop in exit mass flow rate is seen along with the drop in the intensity of mass flow rate fluctuations (shown as error bars in Figure \ref{fig:mass_freq_TR}) as $\zeta$ varies between $0.3 \leq \zeta \leq 0.7$. The drop in mass flow rate is directly proportional to the total pressure loss across the increasing ramp height or the rapid pressure rising in the exit as $\zeta$ increases. The drop in fluctuation intensity is attributed to the rapid rise in the dominant unsteady frequency component and the associated drop in the average exit mass flow rate.
	
	A non-dimensionalized plot of the dominant frequency component ($f$) from the periodically varying exit mass flow rate is plotted for different $\zeta$ in Figure \ref{fig:mass_freq_TR}b. The isolator length($L$) and the stagnation acoustic speed ($a_0$) are used to non-dimensionalize $f$. In general, the quantity $fL/a_0$ is increasing for increasing $\zeta$. However, the trend is not following any unique function. On the other hand, the ratio of the reversed mass running towards the inlet ($m_i$) computed at $[x/L] \sim 0.7$ with respect to the exiting mass towards the outlet ($m_o$) exhibit an exponential trend. The periodically oscillating mass flow rate graphs for different $\zeta$ as shown in Figure \ref{fig:mass_TR}a are used to calculate the reversed flow mass and the variation of the ratio $|m_i/m_o|$ for different $\zeta$ is plotted in Figure \ref{fig:mass_freq_TR}c. Hence, instead of a conventional non-dimensionalization of $f$ using $L$ and $a_0$, the reversed mass ($m_i$) and the average mass flow rate across the inlet ($\bar{\dot{m}}$) are considered for non-dimensionalizing $f$. A typical plot carrying the variation of $[m_if/|\bar{\dot{m}}|]$ for different $\zeta$ is shown in Figure \ref{fig:mass_freq_TR}d, where the exponential rise in the unsteady parameter ($[m_if/|\bar{\dot{m}}|]$) can be seen as $\zeta$ increases. In fact, the underlying dependence of $m_i$ and $\bar{\dot{m}}$ form the foundation for seeking a universal scaling law to predict the dominant frequency observed in a throttling inlet and it is discussed in detail at Sec.\ref{sec:scaling}.
	
	\subsection{On the performance parameters} \label{sec:perf_par}
	
\begin{figure*}
		\centering{\includegraphics[width=1\textwidth]{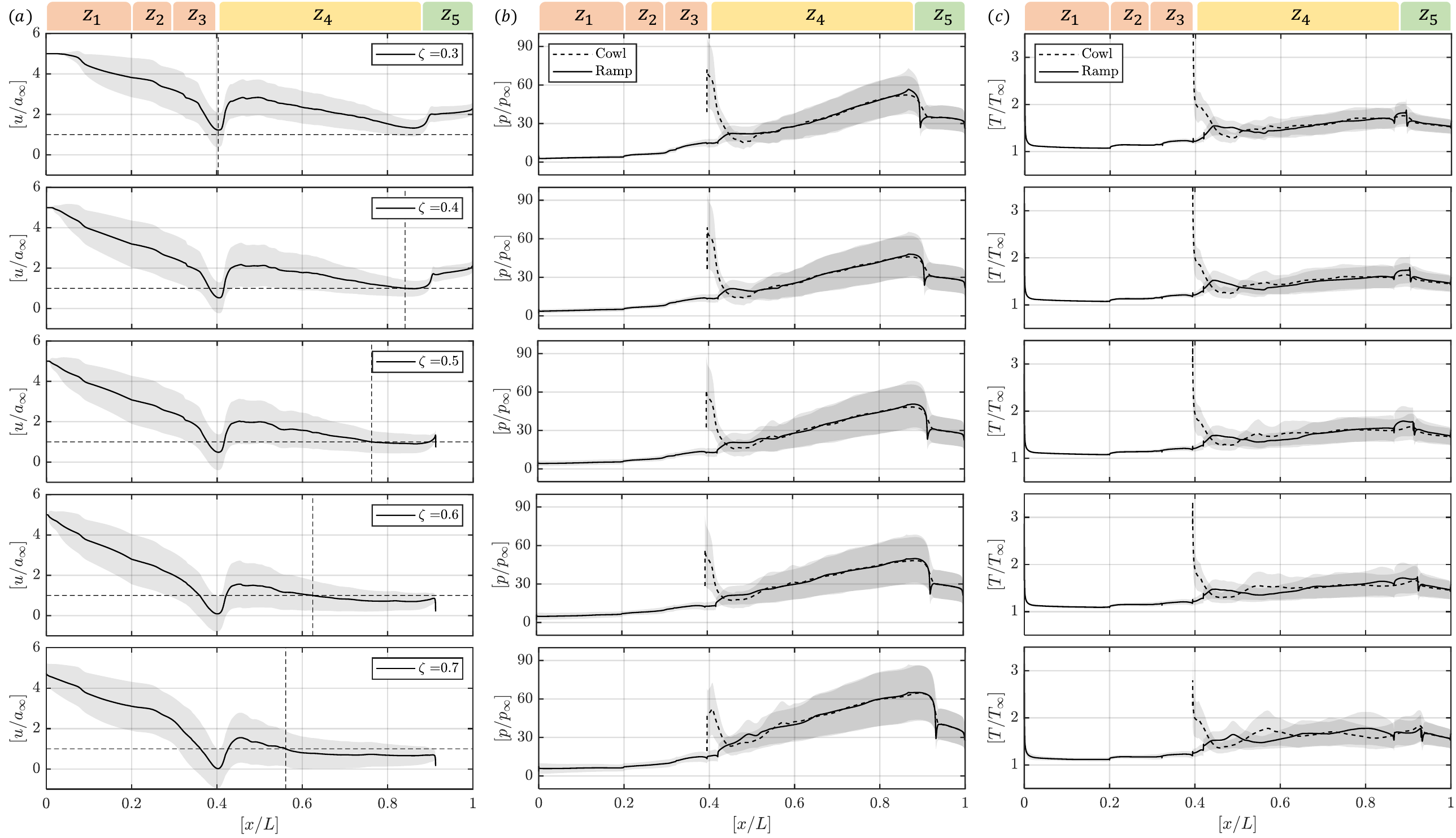}}
		\caption{Plots showing the variation of center-line streamwise velocity ($u$) along the $x$-direction. Sonic point cross-over is marked in the $u$ variations using a dotted black line. Plots showing the variation of center-line static pressure ($p$) and static temperature ($T$) along the $x$-direction on the interior cowl (dotted black line) and ramp (solid black line) wall. Ramp walls are longer than the cowl wall. \sk{The shaded regions mark the standard deviations about the mean.} The zones marked in the figure represent the hypersonic inlet portions: $z_1$-first ramp, $z_2$-second ramp, $z_3$-third ramp, $z_4$-isolator portion, and $z_5$-throttling plug portion.}
		\label{fig:line_data}
	\end{figure*}

	The severity of the unsteadiness during throttling is studied by monitoring the variations in the key fluid parameters like, streamwise velocity ($u$), \sk{static pressure ($p$), static temperature ($T$), total pressure ($p_0$), and total temperature ($T_0$)} between $0.3 \leq \zeta \leq 0.7$. Values of $u$ are monitored by taking a measurement line about the isolator ($h/2$), running from the isolator's exit to the leading edge. Even in the external compression ramp region, the measurement line is maintained at a ramp-wall normal distance of $h/2$, only, as shown in Figure \ref{fig:model}. The time-averaged variations observed in $u$ is plotted about the defined measurement line in Figure \ref{fig:line_data}a. Different parts of the inlet along the measurement line including ramp-1 ($z_1$), ramp-2 ($z_2$), ramp-3 ($z_3$), isolator ($z_4$), and throttling plug ($z_5$) are marked in Figure \ref{fig:line_data}a in accordance with the definitions made in Figure \ref{fig:model}.
	
	At $\zeta=0.3$, the time-averaged streamwise velocity is gradually reduced across the external ramps to a value close to the sonic condition near the isolator's entrance. In the leading edge, as the `inflation' stage is not severe for $\zeta-0.3$, $u$ remains $u_\infty$ for certain $x/L$ at $z_1$. However, as $x/L$ increases, $u$ continues to decrease through ramp generated oblique-shocks. Owing to the presence of separation bubble (see Figure \ref{fig:diff_zeta} for $\zeta=0.3$) in the inlet, $u$ drops suddenly. However, it recovers to a supersonic value immediately upon further expansion across the separation bubble. The value of $u$ further drops gradually in the isolator section by forming multiple-shock systems until the throttling plug, where the value of $u$ reaches closer to the sonic condition. Across the throttling plug, the flow accelerates slightly, especially around the throttle-plug's shoulder, and reaches the outlet at supersonic speed. The shaded region about the time-averaged value of $u$ given in Figure \ref{fig:line_data}a represents the fluctuation intensity as similar to the definition of classical turbulent intensity. The fluctuations are strong during the external compression than in the isolator section. The complete change of shock systems position and the associated changes in fluid velocity and direction, especially from the spillage, cause severe fluctuations in $u$.
	
	As $\zeta > 0.3$, the severity of the `inflation' stage of shock-related unsteadiness increases. At $\zeta=0.7$, the inflation stage enters into the `with-hold' stage, where a bow-shock is standing at the leading edge (see Figure \ref{fig:diff_zeta} for $\zeta=0.7$). It in turn creates severe fluctuations about the mean at $[x/L]=0$ for $\zeta=0.7$. Also, during the `collapse' stage, triple point formation and the associated shedding of strong vortices further reduce the inlet velocity. The combination of reduced inlet velocity along with the supersonic spillage arising during the `inflation' stage renders the time-averaged $u$ to be almost in low-subsonic values at the cowl-lip plane in comparison with the lower $\zeta$ ($\zeta \leq 0.7$). One of the prominent features arising from the inlet's throttling is the upstream movement of the sonic condition (time-averaged) achieved inside the isolator section (marked as dotted-line in Figure \ref{fig:line_data}a). When $\zeta$ changes from 0.4 to 0.7, the location of sonic condition inside the isolator shifts upstream from $[x/L]\sim 0.85$ to $[x/L]\sim 0.55$. The sonic condition's upstream movement ensures the flow to catch-up with the cowl-lip and re-establish the new flow conditions based on the mass imbalance produced from the total pressure loss across the shock. Such rapid upstream movement in a shorter distance increases the frequency of the oscillation. For higher $\zeta$, especially for $\zeta \geq 0.5$, the measurement line is well beneath the flow passage formed by the throttling plug, and hence, no values are shown at $z_5$ region in Figure \ref{fig:line_data}a.
	
	In Figure \ref{fig:line_data}b-c, variation of static pressure ($p$) and temperature ($T$) is plotted on the internal side of the cowl (dotted line) and ramp wall (solid line) (see Figure \ref{fig:model} for locating the internal side of the cowl and ramp wall). Both $p$ and $T$ rises gradually along the external compression ramps through a series of oblique shocks. However, a rapid rise in $p$ and $T$ is evident inside the isolator duct via multiple shock reflections between the cowl and ramp walls. The rapid transition of \sk{undistorted shock train} into a sinuous oscillating shock train to accommodate the upstream propagating wavefronts, which adjust the inlet mass flow rate, leads to the rise in the fluctuation quantity (marked as error-bar in gray color about the mean). The rise in $p$ and $T$ on the cowl and ramp wall as $\zeta$ increases is almost the same towards the isolator's exit (just before the throttling plug). Drop-in, $T$ at the cowl's leading edge for higher $\zeta$ (like $\zeta=0.7$), is due to the severity of the oscillation stages (inflate, with-hold, and collapse), and especially, due to the supersonic spillage. It also forms the reason behind the gradual rise in $p$ seen closer to the isolator entrance as $\zeta$ increases between $0.3 \leq \zeta 0.7$.
	
\begin{figure*}
		\centering{\includegraphics[width=0.8\textwidth]{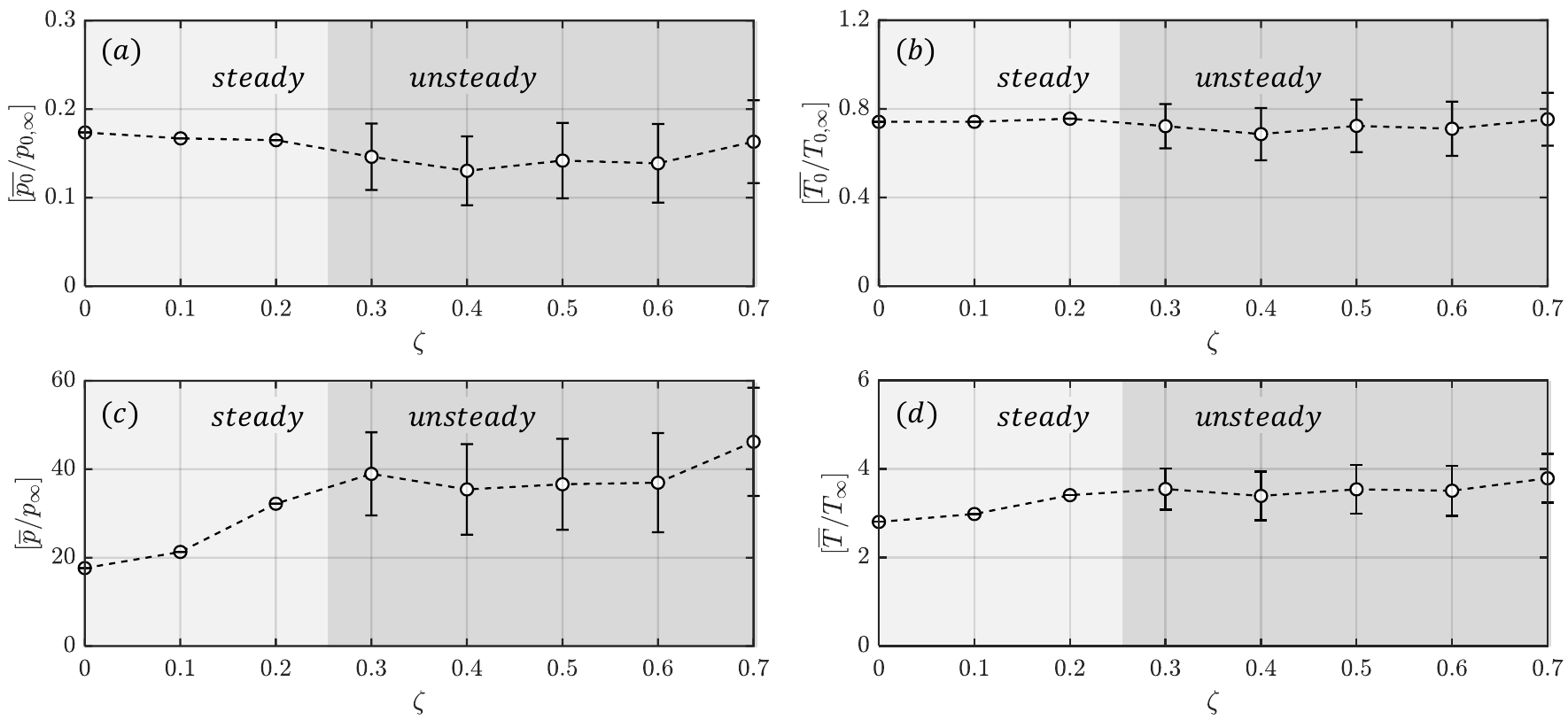}}
		\caption{Plots of different performance parameters seen at the end of the isolator duct in the hypersonic inlet for different throttling ratios ($\zeta$): (a) total pressure loss ($\overline{p_0}/p_{0,\infty}$), (b) total temperature loss ($\overline{T_0}/T_{0,\infty}$), (c) compression ratio ($\overline{p}/p_{\infty}$), and (d) temperature rise ($\overline{T}/T_{\infty}$). \sk{The mean values across the $[y/L]$ profile at the outlet $[x/L]\sim 1$ are plotted through the filled circular markers and the standard deviations are marked as error bars. The error bars exist only for the periodic unsteady events which are prone to exist in the present study between $0.3 \leq \zeta \leq 0.7$.}}
		\label{fig:perf_par}
	\end{figure*}

	The rise in static quantities like $p$ and $T$ and the losses in the total quantities like $p_0$ and $T_0$ form the basis for estimating the inlet's thermodynamic performance. The average of these quantities is considered at the isolator's exit, or the pressure outlet boundary as $\zeta$ varies to comment on the thermodynamic performance. A typical plot in Figure \ref{fig:perf_par} shows the quantities' variation along with the fluctuations (as error-bar about the mean arising due to the periodic variations). Figure \ref{fig:perf_par}a-b show the variations of $p_0$ and $T_0$. As $\zeta$ increases, the loss in $[\overline{p_0}/p_{0,\infty}]$ is very minimal and there is almost negligible loss in $[\overline{T_0}/T_{0,\infty}]$ until $\zeta=0.2$. The falling of throttling conditions well above the Kantrowitz limit in Figure \ref{fig:kant_limit} where the flow is steady might be one of the reasons. The moment $\zeta$ enters into the no self-start regime of Figure \ref{fig:kant_limit}, the losses are severe. A maximum drop of 33.6\% in $[\overline{p_0}/p_{0,\infty}]$ is seen between $\zeta=0$ and $\zeta=0.4$. Similarly, a reduction of 8\% in $[\overline{T_0}/T_{0,\infty}]$ is observed between $\zeta=0$ and $\zeta=0.4$. At higher $\zeta$ ($\zeta$=0.5 to 0.7), the losses are decreasing due to the presence of minimal mass flow rate and the associated formation of several weak oblique shocks at the exit of the isolator. One such instance could be seen in the instantaneous contour plot shown in Figure \ref{fig:comb_cycle}b at $[ft]=2.70$ for $\zeta=0.7$. 	
	
	In Figure \ref{fig:perf_par}c-d, the variations in $p$ and $T$ are given for different $\zeta$. The rise in $p$ is, in general, called compression. As $\zeta$ increases in the steady flow region ($0 \leq \zeta \leq 0.2$), compression drastically increases. In the unsteady regime ($0.3 \leq \zeta \leq 0.7$), the changes in compression are fairly constant, except at $\zeta=0.7$. The deviance is explained by the weak oblique shock formation and lesser mass flow rate into the isolator, as told in the previous paragraph. A large rise in compression of about 1.62 times is seen between $\zeta=0$ and $\zeta=0.7$. On the other hand, the changes in $T$ for different $\zeta$ is gradual. A maximum rise in $T$ of about 36\% is seen between $\zeta=0$ and $\zeta=0.7$. In all the cases represented in Figure \ref{fig:perf_par} for $0.3 \leq \zeta \leq 0.7$, the fluctuations are also fairly constant about the mean. The reason could be attributed to the integral approach in calculating the static and total quantities measured at the pressure outlet boundary. Asides, the plotted mean quantities in Figure \ref{fig:perf_par} shed valuable information on the gas dynamics inside the isolator.	
	\subsection{On the spectral signature of throttling}\label{sec:spectral_sign}
\begin{figure*}
		\centering{\includegraphics[width=1\textwidth]{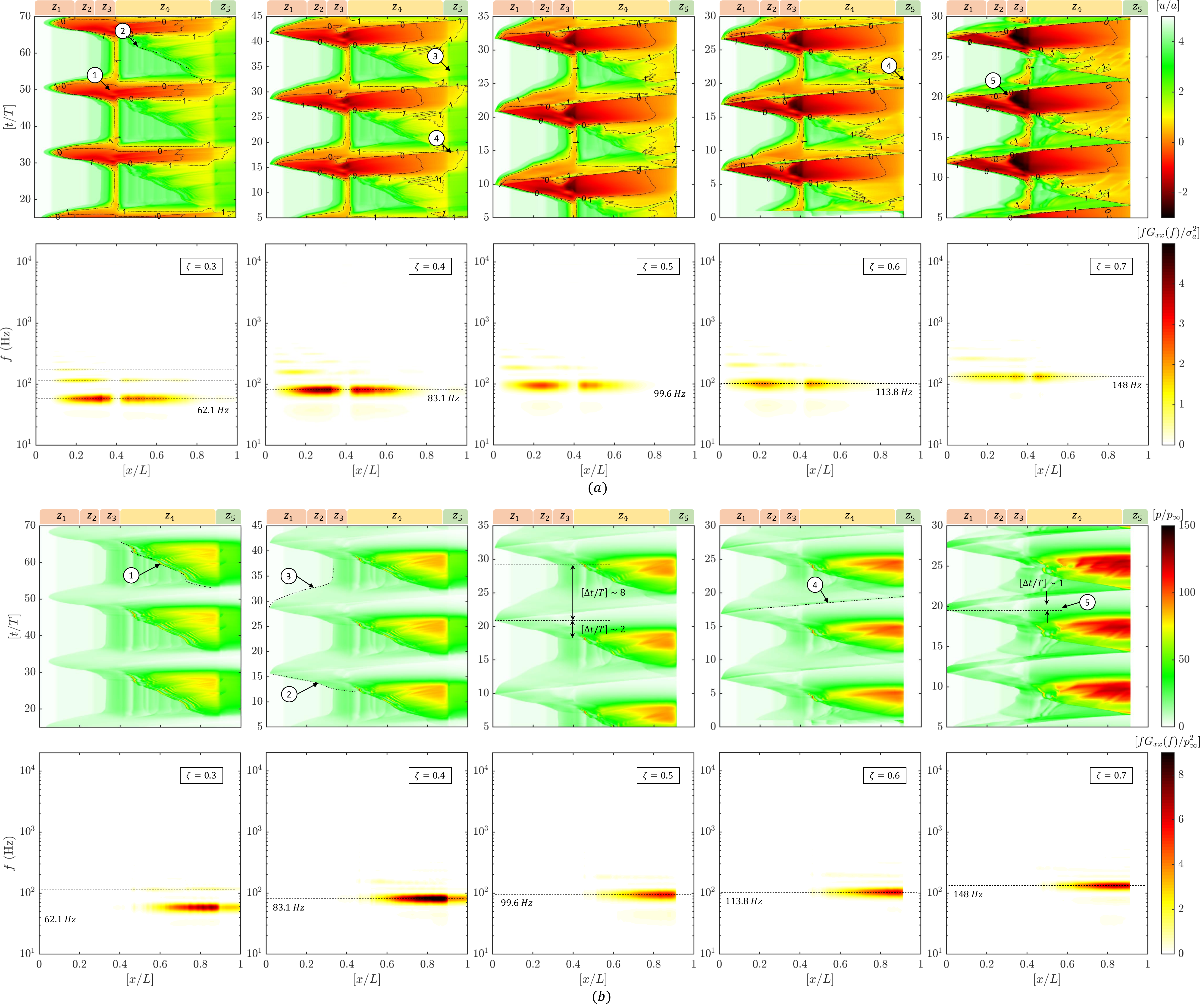}}
		\caption{Contour plots constructed by stacking the center-line profile along the $x$-direction at different instants of $t$ for (a) streamwise velocity ($u$) and (b) static pressure ($p$). Below the $x-t$ contour plot of the aforementioned flow variables, the contour plot obtained by taking the power spectral density at each $[x/L]$ is given. A typical deviation given by $\sigma_a=0.1 a_\infty$ is used for normalizing the power spectral density of $u$, whereas the freestream static pressure ($p_\infty$) is used for normalizing the power spectral density of $p$. Key flow features in (a): 1. reversed mass flow (subsonic spillage), 2. trace of sinuous pseudo-shock movement, 3. supersonic flow at the exit, 4. choked flow at the exit, 5. supersonic spillage. Key flow features in (b): 1. trace of pseudo-shock upstream movement, 2. trace of shock during `inflation' stage, 3. trace of shock during `collapse' stage, 4. trace of downstream movement of shock from the leading edge, 5. short duration of `with-hold' stage at higher $\zeta$.}
		\label{fig:xt_plots}
	\end{figure*}

	The dynamics of throttling can be better understood by plotting the variations of key flow quantities across the isolator's length at different time instants. Performing fast Fourier transform (FFT) or taking power spectral density of such quantities at every spatial point along the inlet's length shed valuable information on throttling's spectral signature. In the present numerical analysis, the spatio-temporal variations of $u$ and $p$ (as they are the quantities showing dominant variations in Figure \ref{fig:line_data}) are taken along the measurement line at $h/2$ for different $\zeta$. Stacking the variations of $u$ and $p$ along the measurement line about the $x$-axis ($x/L$) for each of the time-step ($t/T$, non-dimensionalized with respect to the reference time of $T$=1 ms) produce the necessary $x-t$ contour plot, where the contours represent the non-dimensionalized values of $u$ and $p$. Power spectral density analysis of the constructed $x-t$ contour plot at every $[x/L]$ for the simulated time of $[t/T]$ provides a spectral contour plot. These plots show the presence of different frequency components ($f$) at every $[x/L]$, where the contours show the power ($fG_{xx}$) contained in each of the spectra. A series of such contour plots are generated for different $\zeta$ in Figure \ref{fig:xt_plots} between $0.3 \leq \zeta \leq 0.7$ where the flow is unsteady. The region closer to the throttling plug is not plotted for higher $\zeta$ as the measurement line pass through the throttling ramp instead of the flow passage.	
	
	In Figure \ref{fig:xt_plots}a, $x-t$ and spectral contour plots of $u$ for different $\zeta$ are shown. The increasing severity of the unsteady oscillation stages are seen as $\zeta$ varies from 0.3 to 0.7 in the $x-t$ plots. At $\zeta=0.3$, the `inflation' and `collapse' stages are only present. The shock systems are not even pushed back to the leading edge, and it is displaced to a maximum upstream location closer to $[x/L]=0.1$. The shock systems are pushed and disgorged to the leading edge only after $\zeta \geq 0.5$. At $\zeta=0.7$, the `with-hold' stage is evident by the presence of leading-edge curved shock system sustaining for a significant period of $[\Delta t/T]\sim 1$. Similarly, the period of subsonic spillage at $\zeta=0.3$ is seen in the $x-t$ plot as pale red at $x/L \sim 0.4$, where $[u/a]\sim -1$ (negative sign represents the flow reversal). The temporal regime of subsonic spillage occupies a significantly larger time period of about $[\Delta t/T] \sim 5$ in comparison with the supersonic spillage of $[u/a]\sim -3$ (marked as dark red at $x/L \sim 0.4$) for a time period of $[\Delta t/T]\sim 2.5$ at $\zeta=0.7$. The time required for the upstream wave propagation through the sinuous pseudo-shocks is also higher for lower $\zeta$ and vice versa for higher $\zeta$. The upstream wavefronts are seen as wiggly left-running wavefronts with decreasing slope traveling from the isolator exit to the entrance in the $x-t$ plots. The corresponding spectral contour plot given below the $x-t$ contour plot provides information on the spectral signature. The fluctuations in $u$ are severe closer to the isolator entrance and on the external compression ramp. The spectra's power intensifies for $\zeta=0.4$ in comparison with other $\zeta$, primarily due to the longer duration of mixed spillage observed closer to the inlet. However, a general trend of increasing frequency is evident from the spectral contour plot. The power of $u$ fluctuations on the external ramp for $\zeta=0.7$ is minimal due to the quicker `collapse' of the shock systems.
	
	In Figure \ref{fig:xt_plots}b, the $x-t$ and spectral contour plots are shown for $p$ at different $\zeta$. Like in the previous plots of $u$, the extent of shock motion is seen clearly between the isolator entrance and the leading edge as distinct green lines. Inside the isolator, a general trend of rapid compression is seen closer to the isolator's exit. The right running shock (downstream movement) decreases positive slope from the leading edge until the isolator's exit is common and gets intensified as $ \zeta $ increases from 0.5 to 0.7. The left running sinuous pseudo shock cells (wiggly curves having a negative slope-upstream movement) are visible in all the cases, and the decreasing slope is well aligned with the `inflation' stage of the shock systems in the external ramps. These findings are consistent with the mass flow rate curves at the exit and the time-averaged curves along the inlet, as shown in Figure \ref{fig:mass_TR} and Figure \ref{fig:line_data}. The temporal extent of the compression zones closer to the exit of the isolator is longer for $\zeta=0.3$ ($\Delta t/T \sim 10$, yellow region in Figure \ref{fig:xt_plots}b) than for $\zeta=0.7$ ($\Delta t/T \sim 3$, red region in Figure \ref{fig:xt_plots}b). The spectral contour plots given below the $x-t$ plots reveal the presence of the dominant frequency. The dominant frequency originates closer to the exit of the isolator, and it increases as $\zeta$ increases. The presence of power spectra closer to the exit confirm the origin of upstream disturbance propagation due to backpressure variations\cite{Chen2017}. The dominant spectra from the fluctuations of $u$ near the external ramp and the fluctuations of $p$ closer to the isolator exit are consistent with the findings from the line data given in Figure \ref{fig:line_data}.
	
	\section{Scaling analysis and the prevalence of dimensionality influence} \label{sec:scaling}
	
	As described at the end of Sec.\ref{sec:unsteady_events}, finding the exact moment of unsteadiness and the dominant unsteady frequency component for a considered hypersonic inlet has significant influence in designing the engine airframe and integration procedures for a sustainable hypersonic flight. Unsteady two-dimensional numerical studies are simple in comparison to the three-dimensional counterpart as it saves computational space. However, running simulations for every throttling condition adds further complexity, which could be avoided if a semi-empirical relation is available to estimate the frequency. In the present section, based on the gathered knowledge on the hypersonic inlet-isolator flow physics, a simple scaling analysis is performed using appropriate variables. A semi-empirical relation is formulated to calculate the frequency based on the available design conditions.
	
	\begin{figure}
		\centering{\includegraphics[width=\columnwidth]{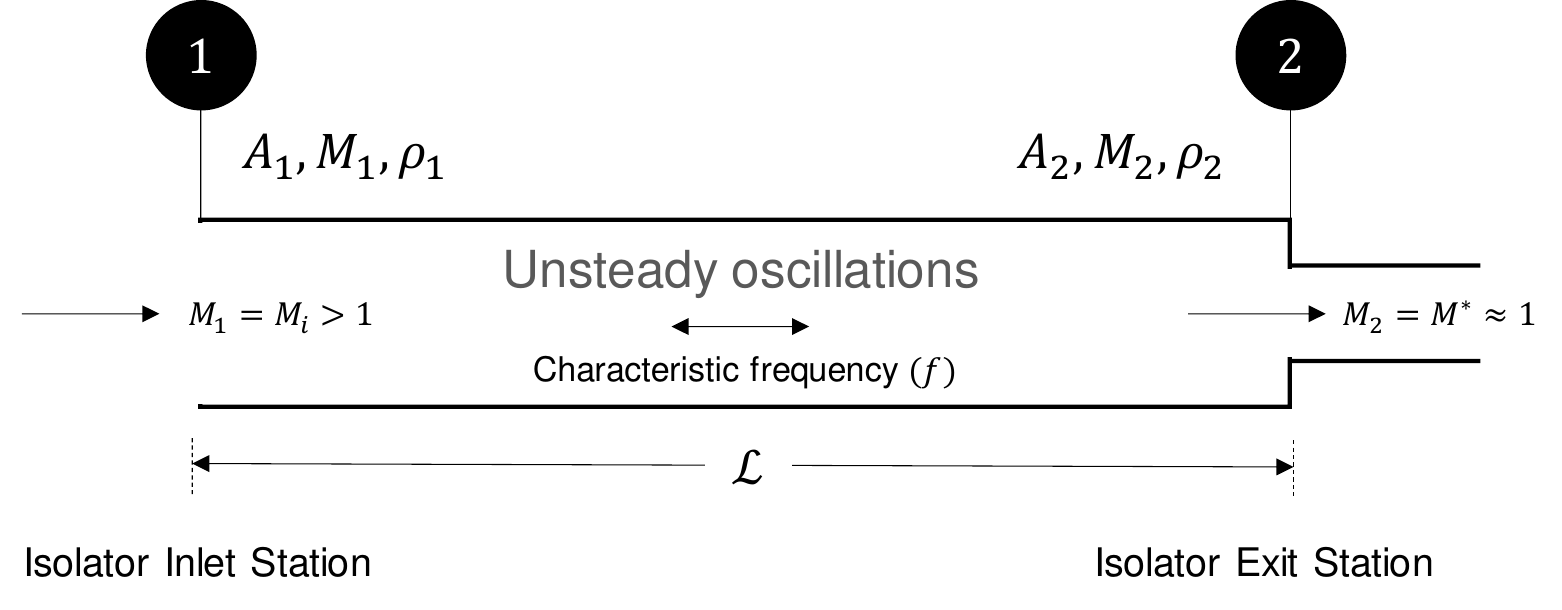}}
		\caption{A two-dimensional sketch showing the key flow and geometrical features to derive the scaling law for the characteristic frequency of unsteadiness ($f$) and throttling ratio ($\zeta$) at different inlet ($M_i$) and freestream ($M_\infty$) Mach number in a typical isolator duct of the hypersonic inlet.}
		\label{fig:scaling_figure}
	\end{figure}
	
	Consider a two-dimensional rectangular duct that represents the isolator potion of the hypersonic inlet, as shown in Figure \ref{fig:scaling_figure}. The isolator duct as two stations: a. inlet (marked as station 1) and b. outlet (marked as station 2). The duct has a constant area ($A_1$) except at the outlet, where the area is $A_2 < A_1$. The dominant unsteady frequency ($f$) contributes to the mass influx and efflux across the isolator duct. It is similar to the identification of influx mass ($m_i$) and efflux mass ($m_o$) as shown in Figure \ref{fig:mass_freq_TR}c-d. The mass flow rate could be scaled using a parameter called $\beta$ using the net mass occupied inside the isolator ($\rho_1A_1\mathcal{L}f$) duct based on the inlet conditions, and the allowed mass flow rate across the outlet ($\rho_2A_2u_2$) as,
	\begin{linenomath}
	\begin{equation}
	\beta = \frac{\rho_1 A_1 L f}{\rho_2 A_2 u_2}=\left(\frac{\rho_1}{\rho_2}\right) \left(\frac{A_1}{A_2}\right) \left(\frac{f\mathcal{L}}{u_2}\right),
	\label{eq-s1}
	\end{equation}
	\end{linenomath}
	where $\rho$, $u$, $A$, and $\mathcal{L}$ are the density, velocity, cross-sectional area and length of the isolator at respective stations. Equation \ref{eq-s1} is modified using Equation \ref{eq:tr} and rewriting $u$ \sk{in terms of} $M$ and $a$ as,
	\begin{linenomath}
	\begin{equation}
	\beta = \left(\frac{\rho_1}{\rho_2}\right) \left(1-\zeta\right)^{-1} \left(\frac{f\mathcal{L}}{M_2a_2}\right).
	\label{eq-s2}
	\end{equation}
	\end{linenomath}
	
	For simplicity, let us define the parameter $\Gamma$ in the isentropic relation\cite{Zucrow1976}, which is a function of $M$ alone (for air being the ideal gas with $\gamma=1.4$) as,
	\begin{linenomath}
	\begin{equation}
	\Gamma(M) = 1 + \frac{\gamma-1}{2}M^2.
	\label{eq-s3}
	\end{equation}
	\end{linenomath}
	
	The term $a_2$, $\rho_1$, and $\rho_2$ in Equation \ref{eq-s2} can be further rewritten using the one-dimensional isentropic flow relations\cite{Zucrow1976} as,
	\begin{linenomath}
	\begin{align}
		\label{eq-s4}
		a_2 &= \sqrt{\gamma R T_2} = \sqrt{\frac{\gamma R T_0} {1 + \frac{\gamma-1}{2}M_2^2}} = \frac {a_0} {\Gamma(M_2)},\\
		\label{eq-s5}
		\rho_2 &= \rho_0 \left(1 + \frac{\gamma-1}{2}M_2^2\right)^{\frac{-1}{\gamma-1}}=\rho_0\Gamma(M_2)^{\frac{-1}{\gamma-1}},\\
		\label{eq-s6}
		\rho_1 &= \rho_0 \left(1 + \frac{\gamma-1}{2}M_1^2\right)^{\frac{-1}{\gamma-1}}=\rho_0\Gamma(M_1)^{\frac{-1}{\gamma-1}}.
	\end{align}
	\end{linenomath}
	
	In the isolator duct, two sectional areas are available, and they can be correlated to the throttling ratio using Equation \ref{eq:tr}. The larger drop in $p_0$ across the throttling ramp associated with the area reduction in the simulation introduces a wavefront to travel upstream to readjust the mass flow rate in the inlet, as discussed in Sec.\ref{sec:unsteady_events} and Sec.\ref{sec:spectral_sign}. Thus, in reality, the flow approaches to a value closer to choking condition at the exit during unsteady throttling. Hence, assuming $A_2 \rightarrow A^*$ well before the initiation of the unsteady events, $M_2\rightarrow M^*\approx 1$. Substituting the values from Equation \ref{eq-s4}-\ref{eq-s6}, and $M_2\rightarrow M^*=1$ in to Equation \ref{eq-s2},
	\begin{linenomath}
	\begin{align}
		\label{eq-s10}
		\beta &= \Gamma(1)\left[\frac{\Gamma(M_1)}{\Gamma(1)}\right]^\frac{-1}{\gamma-1}\left[\frac{\mathcal{L}f}{a_0\left(1-\zeta\right)}\right] = \frac{\Pi(M_1) \mathcal{L}f}{a_0\left(1-\zeta\right)},
	\end{align}
	\end{linenomath}
	where $\Pi(M_1)=1/\Gamma(M_1)^{1/(\gamma-1)}$. The approximation of $M_2\sim M^*=1$ simplified the complexity in Equation \ref{eq-s2}, and renders the parameter $\beta$ being the function only known variables to us like the inlet Mach number ($M_i=M_1$), the isolator length ($\mathcal{L}$), the throttling ratio ($\zeta$), and the dominant frequency ($f$).	
	\begin{table*}
		\caption{\sk{Tabulation of geometrical and flow parameters required to calculate the scaling terms $\beta$ and $\alpha$ as given in Equation \ref{eq-s10} and \ref{eq-s11} from both the existing literature\footnote{A typical value for a geometrical/flow parameter is approximated from the given schematic/flow conditions, if an accurate value is not explicitly stated in the original paper. $\mathcal{L}$ is approximated as the horizontal distance between the cowl-lip and the throttle-plug's minimum area. $M_i$ is calculated at the cowl-lip entrance plane.} and the present numerical studies.}}
		\label{tab:lit_par}
		\begin{ruledtabular}
			\begin{tabular}{@{}lccccccc@{}}    
				Cases &	$M_i$ & $M_\infty$ & $\mathcal{L}$ (m) & $a_0$ (m/s) & $f$ (Hz)\footnote{Only the first dominant frequency is mentioned for the corresponding $\zeta$.} & $\zeta$\\ 
				\midrule
				Li \etal \cite{Li2013} (Exp.) & 4.25 & 5.9 & 0.226 & 570.5 & 222,288,375,400 & 0.4,0.5,0.6,0.7\\
				Wagner \etal \cite{Wagner2010} (Exp.)& 4.31 & 4.9 & 0.333 & 366.9 & 124 & 0.2\\
				Tan \etal \cite{Tan2009} (Exp.)& 3.17 & 4.92 & 0.190 & 482.8 & 190,221,296,337,342 & 0.5,0.58,0.66,0.71,0.85\\
				Kumar \etal \cite{Devaraj2020} (Exp.)& 3.27 & 6.0 & 0.0733 & 410.28 & 950,1100 & 0.55,0.69\\
				Chang \etal \cite{Chang2012c} (Exp.)& 3.48 & 5.0 & 0.25 & 470.1 & 110.3,163.9,177.5 & 0.805,0.874,0.9\\
				Chang \etal \cite{Chang2014a} (Exp.)& 3.75 & 5.0 & 0.25 & 425.22 & 147.8,174.7 & 0.84,0.9\\
				\multirow{3}{*}{Soltani \etal \cite{Soltani2016} (Exp.)\footnote{The results are from the axisymmetric supersonic inlet unlike the others that are from the rectangular hypersonic inlet. These results are considered here to show that the scaling exists irrespective of the cross-section, provided the geometrical parameters are considered properly.}} & 1.51 & 1.8 & 0.5\footnote{\label{Soltani}As the pseudo shock train terminates well before the subsonic flow combustion chamber, the average length of the pseudo shock train during in the isolator is considered as $\mathcal{L}$, as it can only be defined between $M^*\leq M \leq M_i$ inside the duct}& 347.2 & 115.8,131.6,139.7,155.4,164.4 & 0.65,0.675,0.7,0.75,0.8\\
				& 1.68 & 2.0 & 0.5\footnoteref{Soltani} & 347.2 & 90,96,113,127 & 0.675,0.7,0.75,0.8\\
				& 1.84 & 2.2 & 0.5\footnoteref{Soltani} & 347.2 & 80,85,104,120 & 0.675,0.7,0.75,0.8\\
				Rodi \etal \cite{Rodi1996} (Exp.)& 3.47 & 4.03 & 0.135\footnote{At the end of the isolator, the flow is diverted into two chambers where the flow chokes. Hence, the distance between the cowl-lip and the isolator exit is considered as $\mathcal{L}$.} & 343.7 & 300 & 0.34\\
				Present Study (Comp.)& 3.09 & 5.0 & 0.708 & 755.7 & 62.1,83.1,99.6,113.8,148 & 0.3,0.4,0.5,0.6,0.7\\
			\end{tabular}
		\end{ruledtabular}
	\end{table*}
	
	The scaling for the inlet Mach number ($M_i$) to the freestream Mach number ($M_\infty$) along with the different throttling ratios ($\zeta$) could be defined as,
	\begin{linenomath}
	\begin{align}
		\label{eq-s11}
		\alpha = \frac{M_i}{M_\infty}(1-\zeta) = \xi(1-\zeta).
	\end{align}
	\end{linenomath}
	
\begin{figure}
		\centering{\includegraphics[width=0.9\columnwidth]{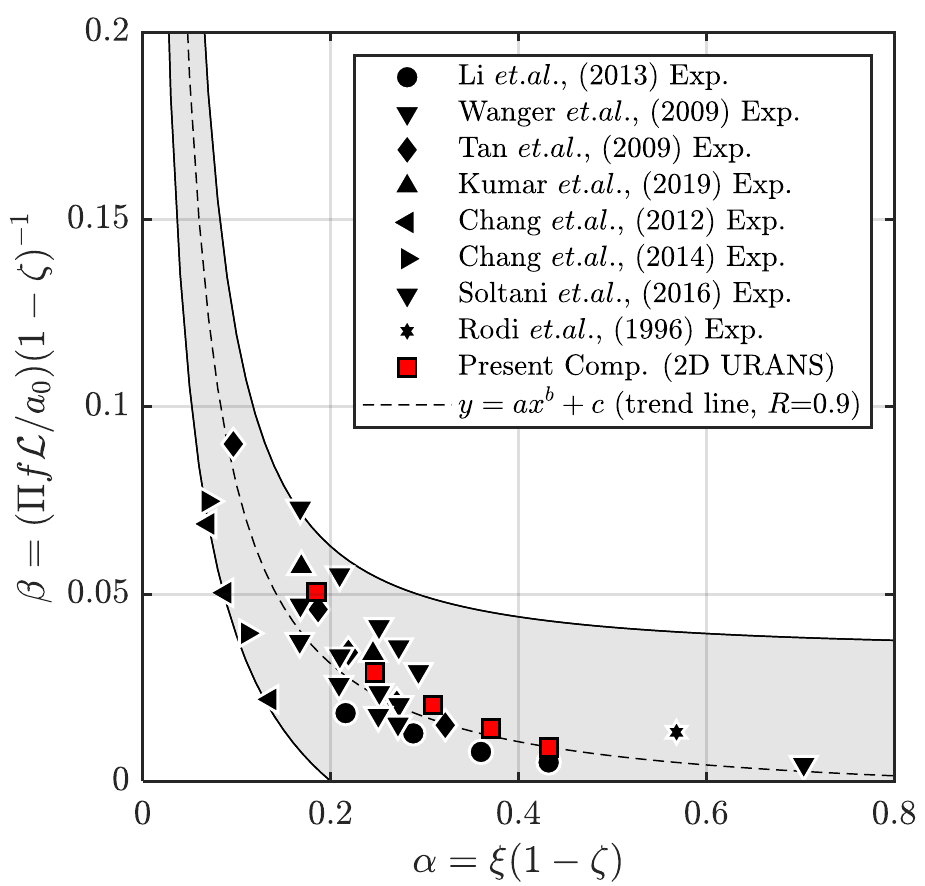}}
		\caption{\sk{Plot showing the variation of scaled throttling parameter ($\alpha=\xi(1-\zeta)$) and scaled dominant unsteady frequency ($\beta=\Pi f\mathcal{L}/a_0(1-\zeta)$). The dotted black line represents the trend-line of type $y=ax^b+c$ with a correlation coefficient of $R\approx 0.9$. The gray shaded region marks the 95\% non-simultaneous observation bound about the trend-line. The filled black markers of different shapes correspond to the experimental data points from the existing literature. The red color filled markers represent the present two-dimensional planar computations and they exhibit a correlation coefficient of $R\approx 0.99$ with the aforementioned trend-line.}}
		\label{fig:scaling_plot}
	\end{figure}

	The parameters given in Equation \ref{eq-s10} and \ref{eq-s11} are computed from the existing literature and also from the present computations, easily. For convenience, the parameters from some of the experiments on hypersonic/supersonic inlet and from the present simulation are tabulated in Table \ref{tab:lit_par}. Some of the vital geometrical and flow parameters are not explicitly given in the literature. An approximated value is selected from the careful analysis of the available sketches and the results. The much needed dominant frequency component $f$ is available in Equation \ref{eq-s10}. We know all the parameters in Equation \ref{eq-s11} at the end of the hypersonic inlet design. Thus by plotting the calculated parameters for $\alpha$ along the $x$-axis and $\beta$ along the $y$-axis as shown in Figure \ref{fig:scaling_plot}, most of the experimental data points are observed to be falling around the trend-line defined as below in Equation \ref{eq-s12} with a correlation value of $R \approx 0.9$,
	\begin{linenomath}
	\begin{align}
		\label{eq-s12}
		\beta = a\alpha^b + c,
	\end{align}
	\end{linenomath}
	where the values of trend-line constants $a$, $b$, and $c$ are 0.0054, -1.2, and -0.006, respectively. The shaded region marks the zone of 95\% non-simultaneous observation bounds about the trend line. With the use of such semi-empirical expression, the experimental data on the unsteady frequency for different $M_i$ can be predicted with ease. For example, in our numerical case, for a throttling ratio of $\zeta=0.7$, $M_i=3.09$, and $M_\infty=5$, the value of $\alpha$ is found to be 0.19. From Figure \ref{fig:scaling_plot} or using Equation \ref{eq-s12}, for $\alpha=0.19$, $\beta$ is estimated to be 0.034. Value of $\beta$ is decomposed for the considered isolator length of $L=0.708$ m, inlet parameter of $\Pi(M_i)=0.07$, and the stagnation acoustic speed of $a_0=755.8$ m/s. The dominant acoustic frequency is thus found to be $f\sim 156$ Hz. The numerically obtained dominant frequency is found to be $f\sim 148$ Hz from Figure \ref{fig:mass_freq_TR}b. The deviation between the predicted $f$ through the semi-empirical formula and the numerical analysis is about 5\%. It has to be noted that the present computations are purely two-dimensional, and in the actual flow, three-dimensional effects\cite{Fisher1986,Reddy1992,Huang2016} pose adverse affects on the dominant spectra to a certain extent. In addition, the semi-empirical relation \sk{has} limitations as it cannot predict the frequency changes precisely during dynamic throttling\cite{Wagner2010,Devaraj2020} or start-unstart hysteresis\cite{Jiao2016,Li2018b}.
	
	As $\zeta$ increases from 0.3 to 0.7, the computed $f$ shown as red-color filled squares in Figure \ref{fig:scaling_plot} deviate from the trend-line, gradually. \sk{Discrepancies between the experimental and computational predictions are known to be in the range of 0.2\%\cite{Abedi2020a} to 10\%\cite{Lu1998}, even though the computations are done only in the two-dimensional sense (like the planar or axisymmetric cases). The difference between the two-dimensional and the three-dimensional simulation also exhibits a maximum deviation of about 2\%\cite{Abedi2020b} when it comes to flow distortion. The gradual deviations in the present computations and the semi-empirical relation could be related to the computational limitations as encountered by the above cases. However, actual experiments are also showing significant scatter about the trend-line in Figure \ref{fig:scaling_plot}. Some of the known issues behind the scatter are due to the real-time effects like the finite width or aspect ratio of the rectangular inlets\cite{Geerts2016}, additional blockage induced by the corner flow separation\cite{Funderburk2016}, and the isolator flow leakage due to flow control like bleeds or improper sealing of the sidewalls\cite{Herrmann2013,SepahiYounsi2018}.}
	
	\sk{Practical inlet design has a finite aspect ratio, and the surrounding wall effects contribute to the generation of other frequency components.} A purely three-dimensional unsteady flow in a simple close-ended rectangular duct is considered to explain the influence of aspect ratio. The dominant frequency from the acoustics theory \cite{Kuttruff2007} based on the dimensions of the rectangular duct is given as,
	\begin{linenomath}
	\begin{align}
		\label{eq-s13}
		f = \frac{a}{2}\sqrt{\left(\frac{i}{l_x}\right)^2+\left(\frac{j}{l_y}\right)^2+\left(\frac{k}{l_z}\right)^2},
	\end{align}
	\end{linenomath}
	where the fundamental or harmonics ($f$) generated by the airflow through it can be represented by the local sound propagation velocity ($a$), mode numbers ($i,j,k$) and length units in the respective dimension ($l_x,l_y,l_z$). As per Equation \ref{eq-s13}, the change in aspect ratio of a rectangular hypersonic inlet affects the dominant $f$ (whose mode numbers are $i=j=k=1$) only by a small factor which is purely a function of $l_y$ and $l_z$. \sk{Besides, the high aspect ratio rectangular isolators are known to have hybrid oblique-normal shocks which modify the unstart behavior notably in comparison to the low aspect ratio ones. The effect of corner flow separation in low aspect ratio isolators exists only at higher frequency and the primary unsteady flow driving low-frequency components are weakly influenced. Boundary layer bleed or leakage produces only a slight shift in the dominant low-frequency component towards the higher/lower end of the spectra. As the present trend-line correlation are drawn from the existing literature of moderate aspect ratio (width/height$\sim$1-6), the scatter is attributed to the above discussed real-time effects. Despite these effects, it has to be emphasized that the proposed semi-empirical relationship can be used at least as a first order estimate in detecting the hypersonic buzz for a wide range of throttling ratio.}
	
	\section{Conclusions} \label{sec:conclusions}
	A simple two-dimensional rectangular hypersonic mixed-compression inlet is considered to understand the unsteady fluid flow observed in the inlet-isolator flow due to throttling by numerical means. The throttling is simulated by placing a wedge-plug of varying heights at the isolator section's exit, thereby simulating different throttling ratios between $0 \leq \zeta \leq 0.7$ in steps of 0.1 as the inlet is exposed to a freestream Mach number of $M_\infty=5.0$. The flow field is achieved computationally by solving the unsteady Reynolds Navier-Stokes (URANS) equations with a $k\omega$-SST turbulence model. The following are the vital conclusions from the present study:
	\begin{enumerate}
		\item {Throttling has negligible effects on the flow quantities, like the exit mass flow rate between $0\leq \zeta \leq0.2$. The flow is observed to be steady. The area contraction achieved by throttling pushes the inlet to enter into the unsteady operation regime, as $\zeta$ increases between $0.3\leq \zeta \leq 0.7$. There is a progressive periodic oscillation observed inside the isolator as $\zeta$ increases. The unsteadiness is of low-frequency type, and the dominant frequency component ($f$) increases with $\zeta$, rapidly.}
		\item {The low-frequency oscillation cycle exhibits three stages. They are classified based on the shock system moving around the external ramp. The three stages are found to be similar to the unsteady shock motion in the spiked bodies at high-speed flows with distinct phases of `inflation,' `with-hold,' and `collapse.' Throttling subsides the occurrence of the `with-hold' stage owing to the subsonic flow spillage from the isolator between $0.3\leq \zeta \leq 0.5$. The presence of all the three stages is evident with increasing severity in flow unsteadiness between $0.6\leq \zeta \leq 0.7$ due to supersonic flow spillage.}
		\item {Key flow parameters like the static ($p,p_0$) and stagnation ($T,T_0$) quantities of pressure and temperature are monitored to identify the influence of $\zeta$ on the performance of the inlet. Loses in the stagnation quantities are severe in the unsteady regime of $0.3\leq \zeta \leq 0.7$. Likewise, in the unsteady regime, the achieved compression or rise in $p$ and $ T $ are at its maximum closer to the isolator's exit.}
		\item {The fluctuations in the flow kinematics and the thermodynamics variables are analyzed by studying the changes in streamwise velocity ($u$) and $p$. They are monitored along a particular measurement line running between the leading edge and the isolator's exit. An $x-t$ contour plot and a spectral contour plot are constructed for different $ \zeta $ about the measurement line, and the inlet's spectral signature is observed. Fluctuations are dominant in the external compression ramp for $u$ and closer to the isolator exit section for $p$. In either case, a rapid rise in $f$ is seen as $\zeta$ varies between $0.3\leq \zeta \leq 0.7$.}
		\item {The variation of dominant low-frequency ($f$) component exhibits to follow an exponential function when scaled with the reversed mass in the isolator as $\zeta$ increases. A universal scaling for the unsteady frequency ($\beta$) is derived using the known parameters of the hypersonic inlet like the freestream and inlet Mach number ($M_\infty$ and $M_i$), isolator length ($\mathcal{L}$), stagnation sound speed ($a_0$), and $\zeta$. A semi-empirical relation is proposed to identify $f$ for any given design conditions of a hypersonic inlet using the existing experimental results in the open literature.}
		\item {Calculated $f$ from the present two-dimensional numerical analysis shows slight deviance from the derived semi-empirical relation, and the reason is attributed to the real-time effects like aspect ratio, corner flow separation, and flow leakage.} 
	\end{enumerate}
	
	\section*{Author's Contributions}
	K.R.S. and S.K.K. have contributed equally to this work. K.R.S. and S.K.K. conceptualized the study, including the design of numerical studies, data analysis, and communication of results. S.K.K., S.J., and R.K. jointly supervised the numerical studies, and took part in the data analysis and the discussion of results. All the authors were involved during the final manuscript preparation.
	
	\section*{Data Availability}
	The data that support the findings of this study are available from the corresponding author upon reasonable request. 
	
	\section*{Acknowledgements}
	\sk{The authors are grateful to the anonymous reviewers for their constructive remarks. The authors acknowledge the invaluable comments of  Dr. Srisha M.V. Rao (IISc, Bengaluru).} The authors thank the Amrita University and Technion for providing the necessary computational facilities to carry-out the simulations and analysis. The first author thanks the Department, faculties, and colleagues for their continuous support and encouragement. The second author thank the Technion for the Post-Doctoral Fellowship offered during his tenure in parts with the Fine Trust, and Prof. Jacob Cohen for his unprecedented encouragement.\\ \\
	\small\textbf{\textsf{REFERENCES}} \bibliography{rajasekar}
\end{document}